# Generation of Ultrafast Magnetic Steps for Coherent Control

G. De Vecchi[1,*], G. Jotzu[1,*], M. Buzzi[1,*], S. Fava[1,*],
T. Gebert[1], M. Fechner[1], A. Kimel[2], A. Cavalleri[1,3]

[1] Max Planck Institute for the Structure and Dynamics of Matter, 22761 Hamburg, Germany

[2] Radboud University, Institute for Molecules and Materials, Nijmegen, The Netherlands

[3] Department of Physics, Clarendon Laboratory, University of Oxford, Oxford OX1 3PU, United Kingdom

e-mail: michele.buzzi@mpsd.mpg.de, gregor.jotzu@mpsd.mpg.de, andrea.cavalleri@mpsd.mpg.de

**A long-standing challenge in ultrafast magnetism and in functional materials research in general, has been the generation of a universal, ultrafast stimulus able to switch between stable magnetic states[1–6]. Solving it would open up many new opportunities for fundamental studies, with potential impact on future data storage technologies. Ideally, step-like magnetic field transients with infinitely fast rise time would serve this purpose. Here, we develop a new approach to generate ultrafast magnetic field steps, based on an ultrafast quench of supercurrents in a superconductor. Magnetic field steps with millitesla amplitude, picosecond risetimes and slew rates approaching 1 GT/s are achieved. We test the potential of this technique by coherently rotating the magnetization in a ferrimagnet. With suitable improvements in the geometry of the device, these magnetic steps can be made both larger and faster, leading to new applications that range from quenches across phase transitions to complete switching of magnetic order parameters.**

* These authors contributed equally to this work

Ultrafast magnetic field steps are unipolar, time varying magnetic field pulses with short risetime and long decay times. Micro coils have been used to generate magnetic field steps with amplitudes of tens of mT and risetimes of hundreds of ps or longer[7–9]. Auston switches can generate faster steps, and are used to produce magnetic pulses with picosecond risetimes and sub-nanosecond decay times[10,11]. In this case, the amplitude of the generated magnetic field is limited by Joule heating in the resistive coplanar waveguides carrying the pulse. Moreover, the spatial profile of the generated magnetic field is generally confined to micron-sized regions (tens of µm) with strong spatial gradients.

Here, we present a new method that overcomes many of these shortcomings, and demonstrate the generation of magnetic field steps that retain the fastest risetimes, long decay times and sizeable amplitudes, while offering new opportunities for tailoring spatial and temporal profiles of the induced magnetic fields.

The physical principle that enables the generation of magnetic steps with picosecond risetimes and "on" times far in excess of a nanosecond (and potentially all the way to milliseconds) is based on a superconducting material with large diamagnetism, quenched with an ultrafast optical pulse. In a zero-field cooled superconductor, the application of a magnetic field generates persistent currents[12,13] to exclude the magnetic flux from the sample volume. If sub-picosecond optical pulses are used to disrupt superconductivity[14–19], a sudden change in the magnetic field surrounding the material generates near field, ultrafast magnetic steps. These pulses are extremely broadband, with frequency content spanning several octaves, from sub-GHz to THz frequencies.

Here, we used a thin film of optimally doped $YBa_2Cu_3O_7$, cooled below $T_c$ and excited with a 100 fs long, 400-nm wavelength optical pulse, which disrupts superconductivity[14–19]. To track the temperature and time evolution of the magnetic field surrounding the superconductor after optical excitation, we used a time-resolved magneto-optical imaging technique[20–22]. We placed a diamagnetic GaP (100) crystal above the superconductor, and the fast Faraday response to the local magnetic field was probed with a time delayed optical pulse with 50 μm spatial resolution. Note that previous realizations of this field detection technique used ferrimagnetic detectors[23–25] that limited the time-resolution to ~100 ps. Here, the use of diamagnetic GaP as a magneto-optic detector allowed for a time resolution better than ~1 ps, at the expense of a smaller signal and longer measurement times (see Supplementary Information S1).

**Thermally Driven Magnetic Field Step**

A 1 mm diameter disc with a well-defined edge was prepared using optical lithography, starting from a 150 nm-thick film of superconducting $YBa_2Cu_3O_7$ grown on $Al_2O_3$. A 75 μm thick, (100)-oriented GaP crystal was placed in close contact with the $YBa_2Cu_3O_7$ film (see Supplementary Information S2 and S3). The two formed a stack with the magneto-optic detector on top and the superconducting device right beneath it (Fig. 1a).

The $YBa_2Cu_3O_7$ sample and GaP detector were subjected to a 2 mT uniform magnetic field applied along the z-direction using a Helmholtz coil pair (see Fig. 1a). A linearly-polarized 800 nm, ultrashort probe pulse was focused to a spot size of ~50 μm FWHM, impinging at near normal incidence on a wedged GaP crystal. The Faraday effect induced a polarization rotation on the beam transmitted through the GaP and reflected from its second surface, yielding a measurement of the vertical component of the local magnetic field, averaged over the volume traversed by the probe in the magneto-optic detector. The external applied magnetic field $H_{app}$ was sinusoidally modulated in time, synchronously to the laser probe pulse train. In this way, signals acquired with $-H_{app}$ were subtracted from those acquired with $+H_{app}$ (see Supplementary Information S4 and S5). This measurement protocol isolated contributions to the polarization rotation having magnetic origin and suppressed potential non-magnetic contributions such as static and pump-induced birefringence.

Because $H_{app}$ was varying in time while the YBa$_2$Cu$_3$O$_7$ was held in its superconducting state, the magnetization in the superconductor originated from exclusion of a time varying magnetic field rather than expulsion of a static magnetic field as in the Meissner effect. This protocol is advantageous compared to the use of static magnetic fields because it minimizes the effect of trapped flux[12] and leads to the largest possible magnetic field steps (see Supplementary Information S6 and S7).

Fig. 1a displays a finite element simulation of the *z*-component of the magnetic field surrounding the sample. We simulate the magnetic field exclusion in YBa$_2$Cu$_3$O$_7$ by modelling it as a medium with virtually infinite conductivity and applying a time dependent external magnetic field (see Supplementary Information S7). The magnetic field is reduced above the sample (blue regions) and is increased near its edges (red regions) as the magnetic flux is kept out from the sample.

The inset on the left-hand-side of Fig. 1b shows a two-dimensional map of the *z*-component of the magnetic field measured when the YBa$_2$Cu$_3$O$_7$ disc is cooled to $T$ = 30 K < $T_c$. As predicted from the simulations, we observed a reduction of the local magnetic field when measuring above the disc (blue area) and a corresponding enhancement near the edge (red area). The amplitude of the measured changes is determined by the geometry of the experiment and by the detection crystal thickness as discussed in Supplementary Information S8. The same measurement was repeated at a temperature $T$ = 300 K > $T_c$. The results are displayed in the right inset of Fig. 1b and show that when the YBa$_2$Cu$_3$O$_7$ disc is in its normal state no spatial dependence of the local magnetic field was detected.

The temperature was then varied between 40 K and 120 K to track the evolution of the magnetic field above the center of the YBa$_2$Cu$_3$O$_7$ disc (see Fig. 1b). As the temperature was swept across $T_c \sim 85$ K, a sharp change in magnetic field was observed, resembling a magnetic field step. The speed at which this magnetic field change can be induced thermally, is limited by how fast the material is heated or cooled through the transition (~ 0.3 K/s) limiting the highest frequency content that can be obtained to less than few Hz.

**Ultrafast Magnetic Field Step**

Picosecond risetimes were achieved by irradiating $YBa_2Cu_3O_7$ with ultrashort optical pulses, which disrupt superconductivity[14-19] on ultrafast time scales. After disruption, we expect the supercurrents shielding the applied magnetic field to disappear on a timescale likely determined by the L/R time constant of the resistive disc (L is the total inductance of the disc and R the resistance in the photoexcited state). As a result, we expect the magnetic field to penetrate back into the sample volume on the same timescales (see Supplementary Information S9).

The geometry of the experiment, shown in Fig. 2a, was the same as that used in the temperature dependent measurements, with the only addition of a quench pulse (400 nm center wavelength, ~100 fs duration, ~0.3 mJ/cm$^2$ fluence), which struck the sample from the side opposite to the GaP detector. Note that the thin $YBa_2Cu_3O_7$ film was opaque to 400 nm radiation, and that the quench beam was shaped as a disc with a flat intensity profile imaged onto the sample to match its size (see Supplementary Information S10). These precautions prevented a direct interaction of the ultraviolet pump with the magneto-optic detector that could give rise to non-linear interactions leading to spurious responses (see Supplementary Information S11 where we show that no magnetic pump-probe signal was detected when the device was kept at 100K, above $T_c$ of the superconductor). Furthermore, the measurements were performed using the differential magnetic field scheme described above, isolating signals only connected to the applied magnetic field $H_{app}$.

Fig. 2b displays the time dependence of the changes in the local magnetic field induced by the quench pulse and measured above the center of the $YBa_2Cu_3O_7$ film at a base temperature $T = 55$ K $< T_c$. As shown in Fig. 1b, at this temperature the $YBa_2Cu_3O_7$ shields the applied magnetic field effectively. As the quench pulse hit the sample, an increase of the magnetic field was detected. This produced an ultrafast magnetic field step of ~0.2 mT amplitude with a risetime of ~1 ps (see inset of Fig. 2b), corresponding to a slew rate of ~200MT/s (further details about this measurement and the fitting procedure are reported in Supplementary Information S9, S12, and S13).

The dynamics of the system are well described by an equivalent L-R circuit when assuming a conductivity of the excited state of 0.1 MS/m, which is lower than, but comparable to, the normal state conductivity of $YBa_2Cu_3O_7$[26,27]. The validity of this model is confirmed by finite-elements simulations, which predict a time constant that scales linearly with the diameter of the disc (more details are described in Supplementary Information S9).

Importantly, after the transient the magnetic field remained constant for many hundreds of picoseconds. Note that in alternative realizations of magnetic field steps, based on Auston switches or by free space THz pulses, the low frequency cutoff is generally limited to tens[10,11,28] and hundreds[29–34] of GHz respectively. In the present case, the lowest frequency achievable is well below 1 GHz, limited only by the time dependence of the externally applied magnetic field (see Supplementary Information S13).

**Coherent control of the Magnetization in a Ferrimagnet**

The broad frequency content of the ultrafast magnetic field steps shown above makes them suitable to control magnetization in a wide variety of magnetic materials that feature magnons and spin-lattice relaxation rates in the sub-GHz to THz frequency range. As a proof of principle, we apply these magnetic field steps to control the orientation of the magnetization in a ferrimagnetic $Lu_{3-x}Bi_xFe_{5-y}Ga_yO_{12}$ garnet[35] (Bi:LIGG in the following). A commercial Bi:LIGG sample with bismuth substitution x ~ 1 and gallium substitution y ~ 1 (see Supplementary Information S2) was used in place of the GaP detector. The geometry of the experiment is shown in Fig. 3a. The in-plane magnetized Bi:LIGG film was ~3 μm thick and grown on a $Gd_3Ga_5O_{12}$ substrate (not shown, see Supplementary Information S2 and S3). Unlike in the measurements shown in Figs. 1 and 2, here we measure the Faraday rotation accumulated by a linearly polarized 800 nm probe pulse traversing the Bi:LIGG layer. In this configuration, the polarization rotation becomes a highly sensitive probe of the magnetization dynamics triggered in the Bi:LIGG[33,36,37]. Similarly to the experiment shown in Fig. 2 a 400 nm quench pulse is used to disrupt superconductivity in the $YBa_2Cu_3O_7$ disc and trigger a magnetic field step that excites the Bi:LIGG sample.

Fig. 3b shows the time dependent changes in the *z*-component of the magnetization triggered by the magnetic field step, measured in correspondence of the center of the disc, with the same pump fluence used in Fig. 2b (~0.3 mJ/cm$^2$). As the magnetic field step excited the Bi:LIGG sample, we observed a pronounced, damped oscillation superimposed to a sudden change in the direction of the magnetization. The data is well fitted with a phenomenological model including a step and a damped oscillatory term (see Supplementary Information S14). The frequency of the oscillations extracted from the fits (~6 GHz) is in good agreement with the ferromagnetic resonance frequency of similar Bi-substituted rare-earth iron garnets[36,37].

The left panel of Fig. 4a shows schematically the time dependence of the magnetic field applied to the Bi:LIGG. $H_{step}$ is the local field generated by the superconducting thin film. At negative time delays it is antiparallel to the external applied field $H_{app}$, due to magnetic flux trapped in the superconductor (see Supplementary Information S13). After the quench pulse hits the YBa$_2$Cu$_3$O$_7$, it promptly becomes parallel to $H_{app}$. $H_A$ lies in the plane of the Bi:LIGG film to account for shape anisotropy. Due to a demagnetizing factor close to unity in our geometry[38], we set $H_A$ equal to the saturation magnetization (~175 kA/m, see below and Supplementary Information S15). While $H_{step}$ varies in time, the anisotropy field $H_A$ is constant. The combination of $H_A$ and $H_{step}$ gives rise to an overall effective magnetic field $H_{eff}$ along which the magnetization points in equilibrium conditions and whose changes determine the dynamics of the magnetization (note that the angles in Fig. 4a are exaggerated for clarity).

The right-hand-side panel of Fig. 4a describes the dynamics of the net magnetization in Bi:LIGG. At negative pump-probe time delays the system is in equilibrium and the magnetization is parallel to the effective field $H_{eff}$. When $H_{step}$ suddenly changes after the disruption of superconductivity in YBa$_2$Cu$_3$O$_7$, the magnetization starts precessing around the new direction given by $H_{eff}$. The z-component of the magnetization oscillates in time at a frequency given by[37,38]: $\omega=\gamma(H_{step} - H_A)$. Neglecting $H_{step}$ (which is much smaller than the anisotropy field) and taking $\gamma/2\pi =·28$ GHz/T[39], we extract a saturation magnetization of 175 kA/m for the measured oscillation frequency of ~6 GHz (Fig. 3b). This value is in good agreement with literature data[36] for similar Bi-substituted rare-earth iron garnets and justifies the assumptions made above.

For comparison, Fig. 4b displays the result of a macro-spin model calculation of the magnetization dynamics following the ultrafast magnetic field step. We set the saturation magnetization of Bi:LIGG to 175 kA/m, we assume the presence of in plane shape anisotropy only (as discussed above) and we vary the damping constant to match our experimental data (see Supplementary Information S15). The results of this calculation are in good agreement with the data shown in Fig. 3b.

**Conclusion**

Disruption of superconductivity in YBa$_2$Cu$_3$O$_7$ discs using ultraviolet laser pulses was used to perturb the magnetic field profile surrounding them, and enabled the generation of magnetic field steps with ultrafast risetimes and super-nanosecond long decay times.

As shown in a proof of principle experiment, ultrafast magnetic field steps open up a new path towards efficient magnetization switching. With the unique properties of our magnetic transient, one could switch the magnetization of a magnetic material with a coercive magnetic field of a mT or less. The size of our device, and therefore the region in which the ultrafast magnetic step is applied, is tailorable. For instance, faster risetimes could be achieved by reducing the size of the superconducting disc down to the micrometer scale, reducing the geometrical inductance of the superconducting disc (see Supplementary Information S9 for more details).

We also foresee possible applications of our technique as a probing tool for quantum materials. For example, a step function with long decay times represents a suitable tool to study persistent currents in (photo-excited) superconductors. Due to its perfect conductivity, a superconductor responds to a step-like magnetic field excitation by creating superconducting currents shielding the magnetic transient for infinitely-long time scales. Because our magnetic field can be switched on at timescales shorter than the lifetime of the transient state, inducing magnetic shielding currents, this time domain technique could be used to study transient superconductors[40–44].

Furthermore, our technique can be applied as a complementary tool to THz time domain spectroscopy, to study low-lying excitations in a large variety of quantum materials given its broadband frequency content, ranging from sub-GHz to 1 THz (see Fig. 5). Because the generated fields are effectively in the near field, our technique may be particularly suitable to study samples with lateral dimensions significantly smaller than the wavelength associated with sub-THz radiation. This is for example the case of novel two-dimensional (anti)ferromagnetic materials[46] which are available only in micrometer-scale sizes.

Finally, we also envision how the amplitude of these magnetic field steps could be made larger using ultrafast demagnetization in ferromagnetic materials[45]

## Acknowledgments

The research leading to these results received funding from the European Research Council under the European Union's Seventh Framework Programme (FP7/2007-2013)/ERC Grant Agreement No. 319286 (QMAC). We acknowledge support from the Deutsche Forschungsgemeinschaft (DFG) via the Cluster of Excellence 'The Hamburg Centre for Ultrafast Imaging' (EXC 1074 – project ID 194651731) and the priority program SFB925 (project ID 170620586). We thank Michael Volkmann, Issam Khayr and Peter Licht for their technical assistance. We are also grateful to Boris Fiedler, Birger Höhling and Toru Matsuyama for their support in the fabrication of the electronic devices used on the measurement setup, to Elena König, Lucie Navratilova, and Guido Meier for help with sample fabrication and characterization.

## Methods

All methods can be found in the Supplementary Information.

## Data Availability

Source data are provided with this paper. All other data that support the plots within this paper and other findings of this study are available from the corresponding authors upon reasonable request.

## Competing Interests

The authors declare no competing interests.

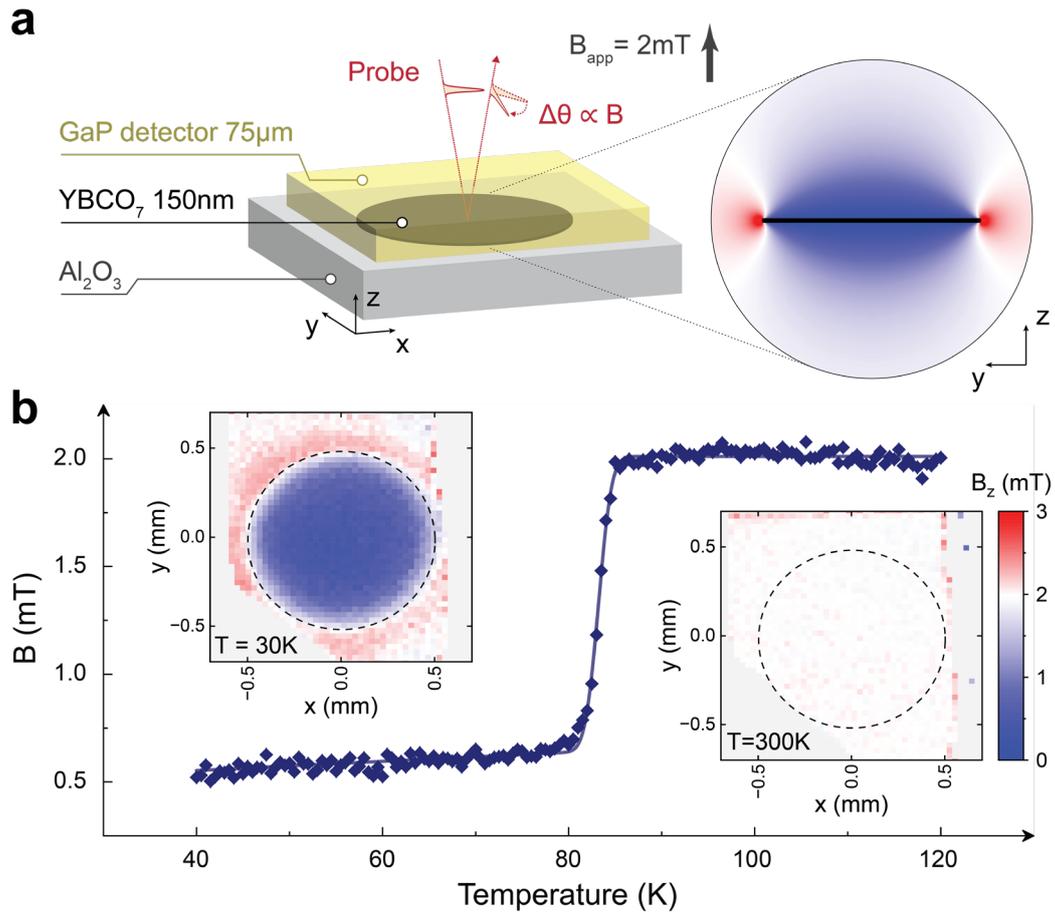

**Figure 1 | Thermally driven magnetic field step. (a)** Sketch of the experimental geometry. A 150 nm thick, 1 mm diameter $YBa_2Cu_3O_7$ disc was grown on an $Al_2O_3$ substrate. The local magnetic field surrounding the $YBa_2Cu_3O_7$ disc is measured by tracking the Faraday polarization rotation of a linearly polarized 800 nm probe pulse, reflected after propagation through a 75 μm thick GaP (100) crystal placed on top of the sample. A 2 mT magnetic field ($B_{app}$) is applied in the z-direction and its polarity is periodically cycled to isolate the magnetic contributions to the polarization rotation. The colour plot (zoomed in view) illustrates the local changes in the *z*-component of the local magnetic field induced by the superconductor below $T_c$ and is the result of a finite element simulation (see Supplementary Information S7 for more details). Blue (red) indicates areas with reduced (enhanced) magnetic field. **(b)** Temperature dependence of the *z*-component of the local magnetic field measured above the centre of the $YBa_2Cu_3O_7$ disc. A clear transition at $T_c$ = 85 K is observed. Note that the measured field exclusion is not complete due to the finite thickness of the GaP detector that leads to averaging in the *z*-direction (see Main Text and Supplementary Information S8 for more details). The left inset shows a two-dimensional map of the *z*-component of the local magnetic field, measured as a function of x and y position at a constant temperature $T$ = 30K < $T_c$. An increase of the magnetic field is measured near the sample edge (red) and a reduction above its center (blue). The dashed black line indicates the outline of the superconducting disc. The right inset shows the same measurement carried out at a temperature $T$ = 300K > $T_c$. No spatial dependence is observed throughout the field of view.

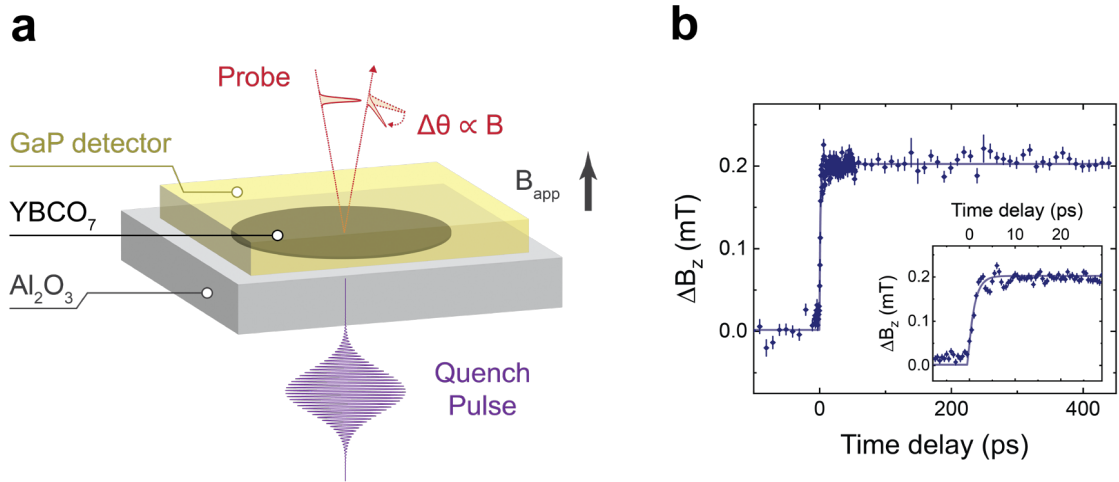

**Figure 2 | Ultrafast Magnetic Field Step. (a)** Sketch of the experimental geometry. The sample-detector assembly is the same as in Fig. 1a. The YBa$_2$Cu$_3$O$_7$ disc is kept at a base temperature $T$ = 55 K < $T_c$ and photo-excited with an ultraviolet laser pulse ($\lambda$ = 400 nm) to disrupt superconductivity. This quench pulse is shaped into a flat-top disc by a shadow mask and imaged by a 4f imaging system onto the YBa$_2$Cu$_3$O$_7$ disc (see Supplementary Information S10). The pump-induced changes in the local magnetic field were quantified using Faraday magnetometry in a GaP (100) detector, as a function of time delay between the probe and quench pulses. A 2 mT magnetic field ($B_{app}$) is applied in the z-direction and its polarity is periodically cycled to isolate the magnetic contributions to the polarization rotation. **(b)** Quench-induced changes in the *z*-component of the local magnetic field, ΔB$_z$, measured above the centre of the disc as a function of pump-probe delay. The diamonds represent experimental datapoints and the solid line is a fit with a single time constant exponential model (see Supplementary Information S9). The inset shows a zoom-in at short time delays around time zero. The value of the measured magnetic field at negative time delays is equal to 0.45 mT due to the dynamics of the trapped flux in the superconductor between consecutive pump pulses (see Supplementary Information S13). The error bars denote the standard error of the mean.

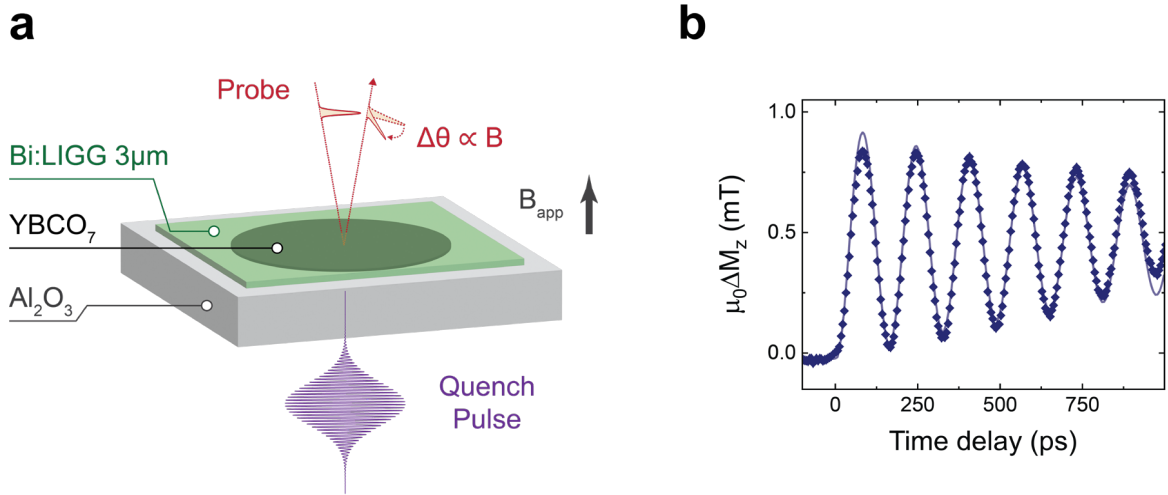

**Figure 3 | Ultrafast Control of Magnetization. (a)** Sketch of the experimental geometry. A thin film of Bi:LIGG replaces the GaP detector on top of the same $YBa_2Cu_3O_7$ disc shown in Fig. 2a. The whole assembly is kept at a base temperature $T$ = 55 K < $T_c$ and the $YBa_2Cu_3O_7$ disc is photo-excited with an ultraviolet laser pulse ($\lambda$ = 400 nm) to disrupt superconductivity. This generates a magnetic field step (see Fig. 2) that is used to trigger the coherent oscillation of a magnon in the neighbouring Bi:LIGG sample. The time evolution of the magnetization following the step excitation was directly quantified using Faraday magnetometry in Bi:LIGG. **(b)** Time dependent changes in the z-component of the Bi:LIGG magnetization $\Delta M_z$, measured above the centre of the superconducting disc as a function of quench-probe delay. The fluence of the quench pulse is ~0.3 mJ/cm$^2$. The value of the z-component of the magnetization at negative time delays is equal to -1.4 mT due to the presence of negative trapped flux before the pump pulse hits the sample (see Fig. 4 and Supplementary Information S13).

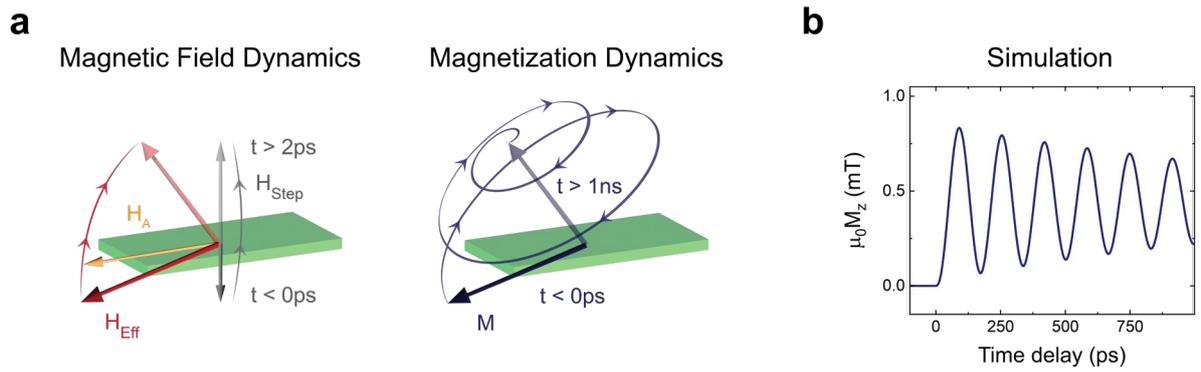

**Figure 4 | Modelling of the observed Magnetization Dynamics in Bi:LIGG. (a)** (left panel) Schematic representation of the dynamics of the effective magnetic field ($H_{Eff}$) inside the Bi:LIGG sample. This is given at any point in time by the sum of a constant anisotropy field accounting for shape anisotropy ($H_A$) and by the time varying magnetic field step ($H_{Step}$) generated by the adjacent superconducting disc. $H_{Step}$ is negative at negative time delays due to the presence of trapped magnetic flux in the superconductor, resulting from the applied magnetic field modulation scheme and the sequence of quench pulses used in the experiment (see Supplementary Information S13). (right panel) Schematic representation of the dynamics of the magnetization (M) in Bi:LIGG. At negative pump-probe time delays, M points along $H_{Eff}$. The sudden change in $H_{Eff}$ resulting from the ultrafast magnetic field step, induces a precessional motion of M around the new direction of the effective field. At longer time scales the magnetization aligns again with the new direction of the effective field. **(b)** Micromagnetic calculations in the macrospin approximation reproducing the experimental conditions using as an input the magnetic field step shown in Fig 2b (see Supplementary Information S15). The plot shows the time evolution of the z-component of M following the excitation step measured above the centre of the YBa$_2$Cu$_3$O$_7$ disc. The calculations are in good agreement with the data reported in Fig 3b.

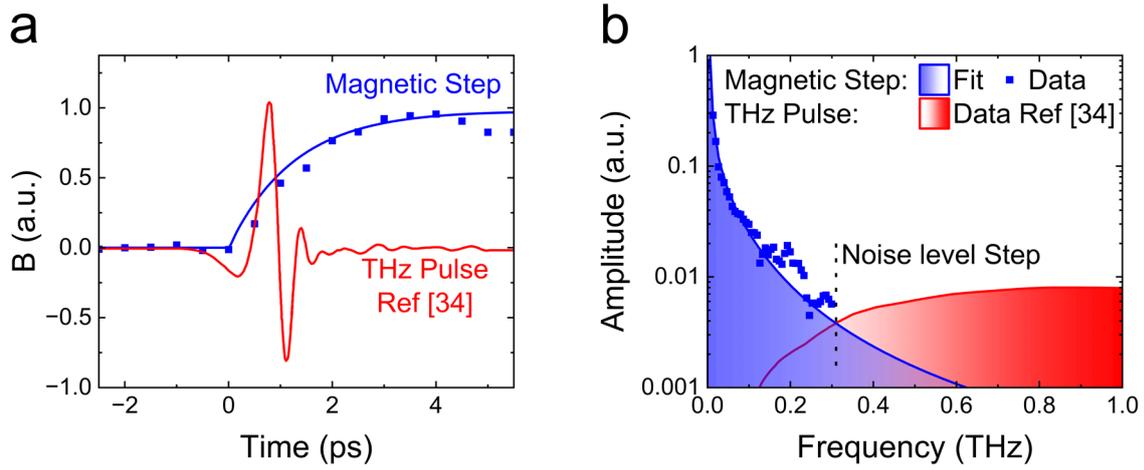

**Figure 5 | A complete set of tools for low-frequency magnetic spectroscopy. (a)** Experimental data (blue squares) and fit (blue solid line) for a magnetic step generated with our technique and amplitude of the measured electric field of a single cycle THz Pulse from reference [34] (red solid line). The two signals are normalized to have the same peak value. **(b)** Fast Fourier Transform of the time domain traces in (a). The experimental data relative to the magnetic step (blue squares) is shown up to 0.3 THz, frequency above which the signal falls below the noise threshold.

# Generation of Ultrafast Magnetic Steps for Coherent Control

G. De Vecchi[1,*], G. Jotzu[1,*], M. Buzzi[1,*], S. Fava[1,*],
T. Gebert[1], M. Fechner[1], A. Kimel[2] and A. Cavalleri[1,3]

[1] Max Planck Institute for the Structure and Dynamics of Matter, 22761 Hamburg, Germany

[2] Radboud University, Institute for Molecules and Materials, Nijmegen, The Netherlands

[3] Department of Physics, Clarendon Laboratory, University of Oxford, Oxford OX1 3PU, United Kingdom

e-mail: michele.buzzi@mpsd.mpg.de, gregor.jotzu@mpsd.mpg.de, andrea.cavalleri@mpsd.mpg.de

## Supplementary Information

# S1. Ultrafast Magnetometry Technique

Ultrafast optical magnetometry relies on the Faraday effect which directly relates the magnetic field applied to a material to the polarization rotation of a linearly polarized beam traversing the medium. This relation is normally reported as:

$$\theta = V \cdot \int_0^L B(z)dz$$

where $\theta$ represents the rotation of the polarization of the input beam, $B(z)$ is the magnitude of the magnetic field along the light propagation direction inside the medium and L is the thickness of the medium. The proportionality constant V is known as the Verdet constant, which is a material dependent constant depending also on other parameters such as the wavelength of the incoming polarized light. This expression highlights the fact that the magnetic field is averaged across the thickness of the material traversed. Therefore, if B varies strongly over distances of order ~L, its spatial dependence is blurred in the accumulated polarization rotation.

In the past, the Faraday effect in ferromagnetic crystals and thin films has been used to image the magnetic properties of superconductors at equilibrium[1,2]. These types of detectors (such as $Bi:R_3Fe_5O_{12}$, EuS, and EuSe) offer very high sensitivity ($V \sim 10^5$ rad·T$^{-1}$·m$^{-1}$) but have limited time resolution, down to 100 ps at best, due to the presence of low lying magnetic excitations (e.g. ferromagnetic resonance) at sub-THz frequencies. On the other hand, diamagnetic II-VI and III-V semiconductors such as ZnSe, ZnTe and GaP have a magneto-optic response featuring Verdet constants that are two to three orders of magnitude smaller than those observed in ferromagnetic materials but have the advantage of not being magnetically ordered and offer significantly better time resolution[3,4]. Furthermore, their Verdet constant is mostly temperature independent ensuring a flat detector response in a broad temperature range.

The measurements shown throughout the manuscript were performed using GaP and $Lu_2Bi_1Fe_4Ga_1O_{12}$ magneto-optic crystals. Their sensitivity to magnetic fields was calibrated before each measurement in the exact experimental conditions then used for the measurement. We measured the amount of angular polarization rotation in a position of GaP or $Lu_2Bi_1Fe_4Ga_1O_{12}$ far from the sample (see Fig. 1 of the Main Text), where the magnetic field was equal to the uniform externally applied magnetic field. In this way we could extract a calibration factor specific of our experimental conditions which allowed us to convert the measured polarization rotation into the value of the local magnetic field in the crystal.

Finally, to have an accurate calibration of the real value of the magnetic field, the current-to-magnetic field constant of the Helmholtz coil pair was independently calibrated in-situ using a Lakeshore 425 gaussmeter. These calibration measurements, after accounting for the magneto-optic crystal thickness, yielded a Verdet constant of ~120 rad·T$^{-1}$·m$^{-1}$ for GaP, in agreement with reported literature values[5], and of ~0.5·10$^5$ rad·T$^{-1}$·m$^{-1}$ for Lu$_2$Bi$_1$Fe$_4$Ga$_1$O$_{12}$, in agreement with the values specified by the company producing the samples (see Supplementary Information S2).

## S2. Sample Characterization

The optimally doped YBa$_2$Cu$_3$O$_7$ thin films were obtained through a commercial supplier (Ceraco GmbH) and grown on r-cut Al$_2$O$_3$ substrates. We chose the S-Type films, with a thickness of approximately 150 nm, critical current density greater than 2 MA/cm$^2$ and specified T$_c$ of 86 K. The superconducting transition temperature of the device after lithography (see Supplementary Information S3) was 85 K (see Fig. 1b of the Main Text). The slight copper excess used by the manufacturer to optimize the superconducting properties of the sample results in small copper-oxide precipitates on the surface of the YBCO film. These defects impact the magnetic properties of the sample as discussed in Supplementary Information S6 and S7.

The ~3μm thick Bi-substituted lutetium iron gallium garnet (Lu$_2$Bi$_1$Fe$_4$Ga$_1$O$_{12}$, Bi:LIGG for short in the following) sample was obtained through a commercial supplier and grown by Liquid-Phase Epitaxy on a (100)-oriented, 500 μm thick Gd$_3$Ga$_5$O$_{12}$ crystal (GGG). The product chosen was the sensor type C. The composition of the Bi:LIGG sample was measured using energy-dispersive X-ray spectroscopy (EDX) yielding the following percentage composition of elements: O 58.98%, Fe 19.43%, Ga 5.99%, Lu 10.28%, Bi 5.32%. This result is in good agreement with the chemical formula reported above.

## S3. Sample preparation and Experimental Geometries

A 150 nm-thick YBa$_2$Cu$_3$O$_7$ film grown on a two side polished Al$_2$O$_3$ substrate was patterned into a disc shape using a laser lithography process based on a AZ1512 photoresist mask. After exposure and development, the sample was wet etched using a 1% H$_3$PO$_4$ solution. After etching, the residual photoresist was removed using acetone and isopropanol. Fig. S3a shows a micrograph of the YBa$_2$Cu$_3$O$_7$ film after patterning. The thin film and GaP (100) detector were then mounted onto an Al$_2$O$_3$ plate that could be fixed directly to the cold finger of the cryostat. Al$_2$O$_3$ is an electrical insulator, and minimizes the shielding of magnetic fields due to eddy currents while ensuring a good cooling power. A 75 μm thick GaP (100) crystal (SurfaceNet GmbH) was used as a detector and put in close contact with the sample (see Fig. S3b). The detector was polished with a wedge angle of ~1.5° to spatially separate the reflections from the front and back surface. This allowed to detect exclusively the reflection from the back surface, which accumulated the Faraday polarization rotation signal as the probe beam propagated across the detector thickness. Additionally, the GaP back surface and the sample were not coplanar, to avoid interference between the reflections from these two surfaces (see Fig. S3c). The gap between detector and the YBa$_2$Cu$_3$O$_7$ disc was ~25 μm. This experimental geometry was used for the measurements shown in Figs. 1 and 2 of the Main Text. For those reported in Fig. 3, the same experimental geometry was used, but GaP detector was directly replaced by the ferrimagnetic Bi-substituted rare earth iron garnet sample. The surface of the sample exposing the ferrimagnetic layer was placed close to the YBCO$_7$ disc, while the wedge in this case was polished on the GGG substrate facing the incoming probe beam.

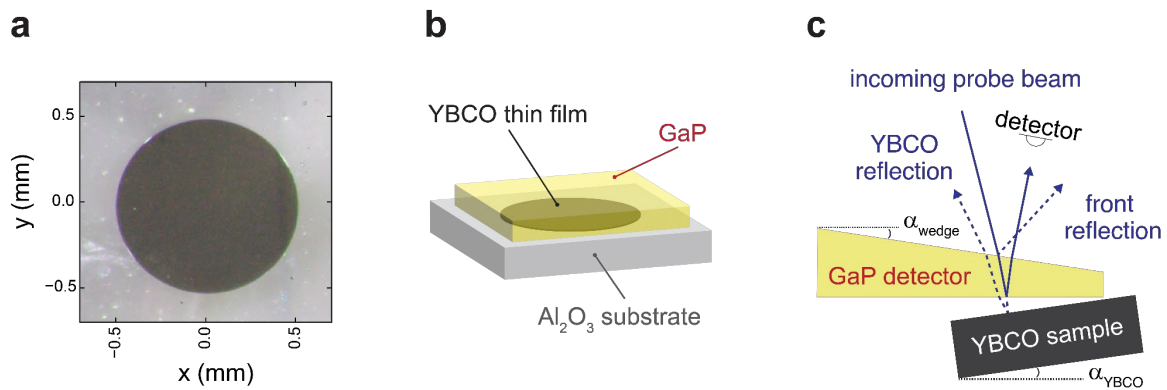

**Figure S3.** | **(a)** Micrograph of the YBa$_2$Cu$_3$O$_7$ thin film patterned into a 1 mm diameter disc shape. **(b)** Sketch of the sample configuration highlighting the positioning of the GaP (100) magneto-optic detector on the YBa$_2$Cu$_3$O$_7$ disc. **(c)** Side view highlighting how undesired reflections are filtered out. The GaP detection crystal is wedged to spatially separate the front and back reflections from each other. The YBCO sample is then mounted with a small tilt with respect to the detector back side. This angle is big enough to filter out the reflection from the YBCO surface, but small enough not to interfere with the measurement. Both angles are exaggerated for clarity, in reality $\alpha_{wedge}$ ~1.5° and $\alpha_{YBCO}$ ~1°.

## S4. Experimental Setup and Data Acquisition

The measurements were performed using the experimental setup sketched in Fig. S4. Ultrashort (100 fs) 800 nm center wavelength laser pulses were produced starting from a commercial Ti:Al$_2$O$_3$ oscillator/amplifier system that produced pulses with energies of 2 mJ at a repetition rate of 900 Hz. These pulses were split using a beamsplitter into two branches. The lowest intensity branch was used after attenuation for probing Faraday rotation in the GaP (100) detector and Bi:LIGG samples. To minimize the noise sources in the measurement, the polarization of the beam was set using a nanoparticle high-extinction ratio linear polarizer.

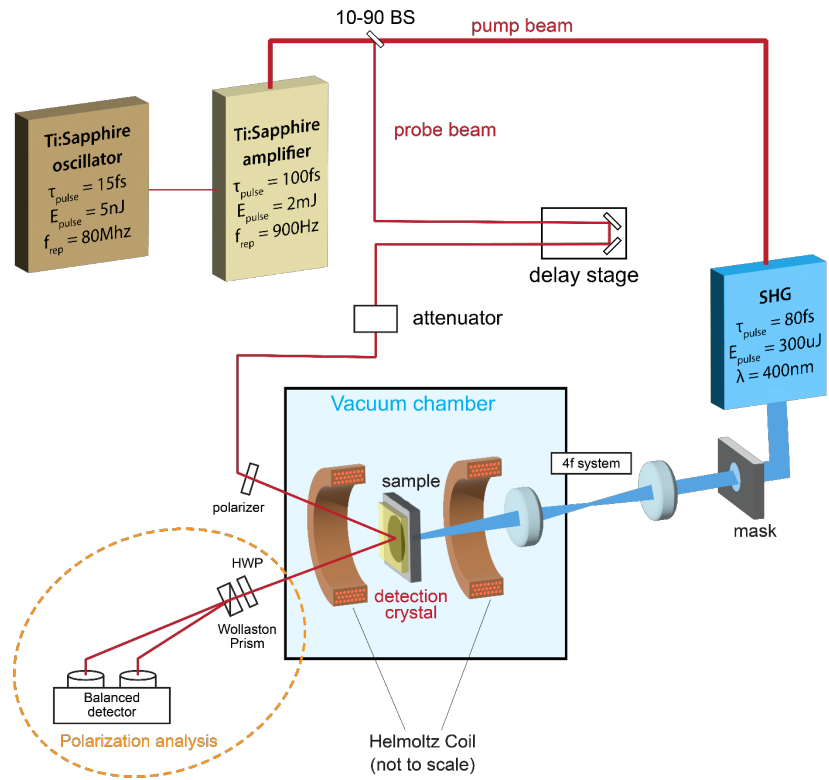

**Figure S4.** | Experimental setup.

As non-normal incidence reflections introduce a phase delay between *s* and *p* polarization, incidence angle fluctuations can give rise to polarization noise. To minimize this, only reflections close to normal incidence were used in the setup and a commercial system using active feedback was used stabilize the laser beam pointing. After traversing and being reflected from the second surface of the Faraday detector, the polarization state of light was analyzed using a half-waveplate, Wollaston prism and balanced photo-diode setup that allowed us to quantify the Faraday effect in the magneto-optic detection crystal. The higher intensity branch was frequency doubled using a β-BaB$_2$O$_4$ crystal to obtain 400 nm pulses that were used to photo-excite the YBa$_2$Cu$_3$O$_7$ thin film sample. A mask, illuminated by these ultraviolet pulses, was imaged onto the back surface of the sample to create a flat top, disc-shaped beam that matched exactly the size of the disc-shaped

YBa$_2$Cu$_3$O$_7$ (see Supplementary Information S10). This, together with the YBa$_2$Cu$_3$O$_7$ being opaque to 400 nm radiation ensured that the GaP detector was not exposed to the pump pulses.

The YBa$_2$Cu$_3$O$_7$ thin film samples were embedded in the detector assembly (Fig. S3) and mounted on the cold finger of a liquid helium cryostat to allow for temperature control. The cryostat was directly placed in a high vacuum chamber. A pair of coils in a Helmholtz configuration generated a magnetic field at the sample position whose polarity could be reversed at a frequency of 450 Hz. Switching the polarity of the magnetic field helped rejecting possible spurious contributions (see Supplementary Information S5) arising from unwanted sources of bi-refringence (e.g. residual strain in the detector crystals or vacuum windows). The highest achievable magnetic field was limited by heat dissipation and was ~3 mT. The sample position was changed using computer controlled linear translation stages that made it possible to reproducibly move the cryostat and the sample inside the vacuum chamber with ~10 µm repeatability.

To obtain differential magnetic field measurements the electrical pulses from the balanced photodetector were digitized using a commercial 8 channel 40MS/s data acquisition card, triggered at the lowest frequency used in the experiment. These pulses, acquired in the time-domain, were then integrated applying a boxcar function to yield the signals from the sum and difference channels of the balanced photodetector for each probe laser pulse (see Fig. S5 in the following section). Since the acquisition of a full pulse sequence required the acquisition of many cycles of the applied magnetic field, the sample clock signal of the data acquisition card was derived using direct digital synthesis from the oscillator repetition rate. In this way drifts in the cavity length and repetition rates of the system did not affect the relative timing of the boxcar functions with respect to the arrival time of the electrical trigger pulse.

## S5. Data Reduction and Analysis

As mentioned in the previous section the polarity of the magnetic field was cycled periodically to yield differential measurements and isolate contribution to the polarization rotation that were only induced by the applied magnetic field. In other words, because the equilibrium and pump-induced magnetic field changes measured with applied field -$H_{app}$ were subtracted from those acquired with applied field +$H_{app}$, the sensitivity of the signal to strain in the magneto-optic detector or polarization noise in the setup was strongly reduced. In the following we discuss this approach in detail.

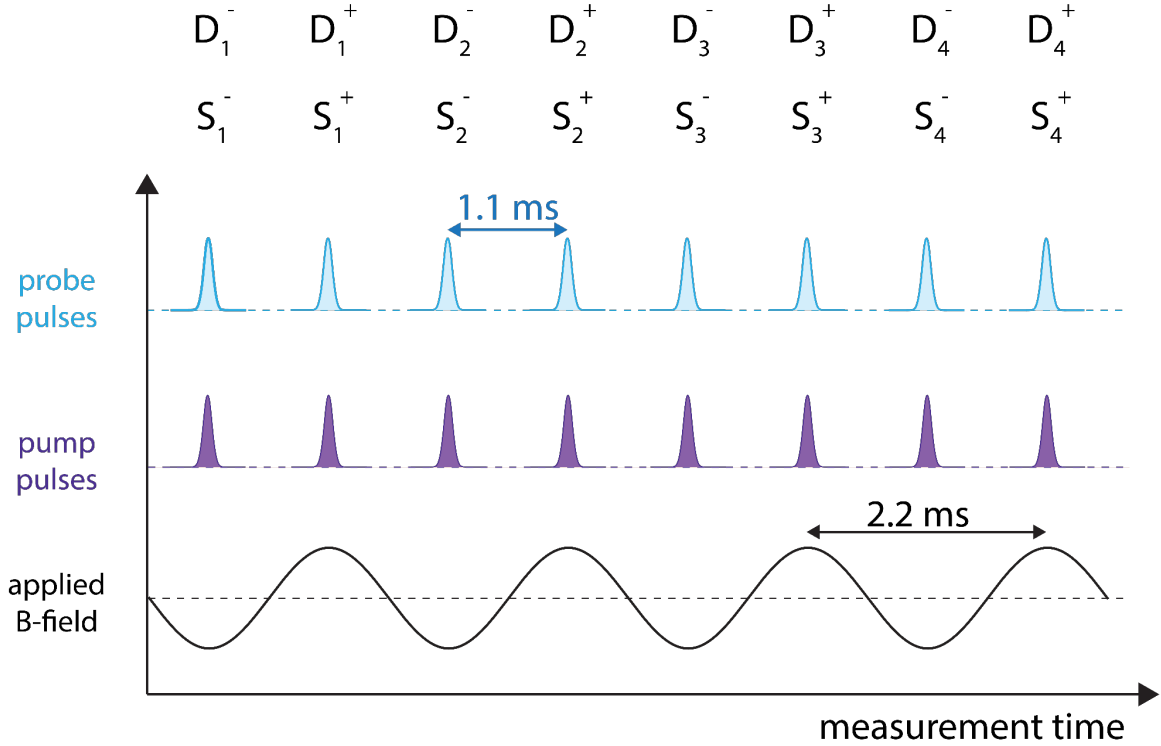

**Figure S5.** | Timing diagram of the acquisition scheme used for the measurements shown in the Main Text. We indicate with D and S the signals from the difference and sum channels of the photodetector respectively.

The magnetic field polarity was cycled following a sinewave at 450 Hz frequency. We chose the highest possible switching frequency to efficiently reject low frequency 1/f noise. A timing diagram of the acquisition scheme is shown in Fig. S5. As mentioned in section S4, we triggered the data acquisition card at a sub-harmonic frequency of the probe repetition rate corresponding to traces comprising 88 pulses (a trace of 8 pulses is shown in Fig. S5 as an example). Only 80 of these 88 pulses were, in the end, acquired due to a fixed dead time for data processing.

We label signals from the difference and sum channels of the photodetector as $D_i^\pm$ and $S_i^\pm$ respectively, to indicate whether they were acquired with positive (+) or negative (-) polarities of the applied magnetic field. The subscript *i* runs over the 80 acquired pulses in the acquisition trace. The polarization rotation within a trace is calculated in the following way:

$$\Delta\vartheta = \frac{\sum_i^{80} D_i^+}{\sum_i^{80} S_i^+} - \frac{\sum_i^{80} D_i^-}{\sum_i^{80} S_i^-}.$$

This quantity yielded the amplitude of the magnetic field, after calibration of the Faraday effect in the GaP (100) detector and in the Bi:LIGG sample (see Supplementary Information S1). To limit computational dead times between acquisitions, we normalize after averaging as indicated in the formula above.

The sequential acquisition of multiple pulses (88) allowed us to minimize the impact of lost pulses during processing time, while keeping the frequency of acquisition as high as possible to reject the effect of low frequency noise when averaging the difference and sum channels within a trace before taking the ratio of the two quantities. To cancel out residual drifts due to possible asymmetries in the applied magnetic field, the phase of the sinusoidal applied magnetic field with respect to the laser pulse train was periodically alternated between 0 and $\pi$.

In the pump probe measurements, each probe signal was acquired with the optical pump on (see Fig. S5). This was convenient to reduce asymmetries due to long lasting heating effect in the superconducting film (see Supplementary Information S13). The magnetic dynamics induced by the pump was therefore measured by changing the value of the probe arrival time with respect to the pump.

## S6. Shielding of Time Varying Magnetic Fields in Superconductors

When cooling a superconductor below $T_c$ in a magnetic field two processes can take place depending on the temporal evolution of the applied field. First, if the applied field is static, we observe the Meissner effect: the expulsion of a static magnetic field from the volume of the superconductor. In a type II thin film superconductor[6,7] with defects[8], like our $YBa_2Cu_3O_7$ sample, this effect can be rather small due to the formation of vortices that trap magnetic flux within the sample volume and reduce its capability of expelling applied magnetic fields (see Supplementary Information S7). Second, if the applied magnetic field varies in time, we observe a magnetic shielding due to the perfect conductivity of the $YBa_2Cu_3O_7$ sample (see Fig. S9.3a). This response is observed in a zero-field cooled experiment[8,9]. In this second case, magnetically induced persistent shielding currents keeps the magnetic flux constant in the superconductor, as prescribed by Faraday-Lenz law, and the magnetic field is completely excluded from the sample volume, even if defects are present. Thus, this second process produces a much bigger magnetic response than the first one and we chose to exploit this to our advantage in our measurement protocol.

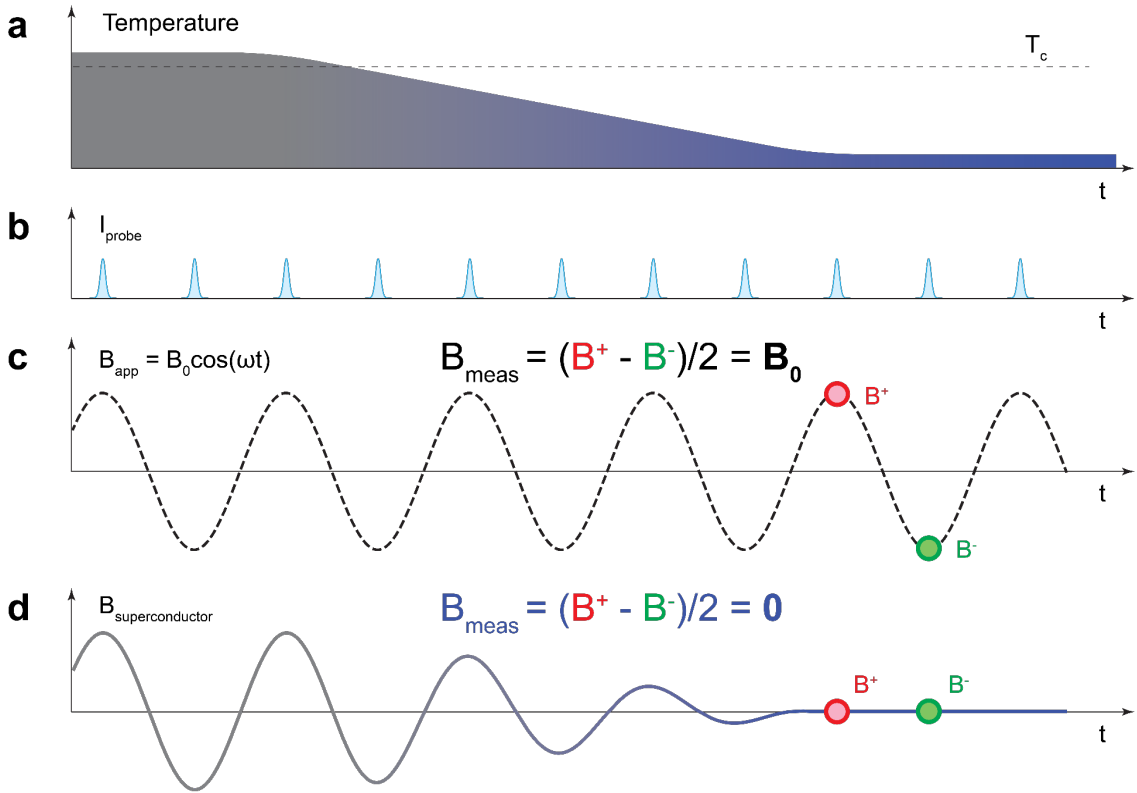

**Figure S6.** | **(a)** Sketch of the temporal profile of the temperature when cooling the sample below $T_c$. **(b)** Arrival time of the probe pulses sampling the magnetic field. **(c)** Externally applied magnetic field, with a sinusoidal time evolution. **(d)** Magnetic shielding of the superconductor when cooling across $T_c$.

Fig. S6 qualitatively explains the equilibrium field dynamics in our experiment, which is very similar to the zero-field cooled regime. Fig. S6a shows the temperature evolution during a hypothetical experiment in which a sample is progressively cooled below $T_c$. Fig. S6b shows the probe pulses used to sample the magnetic field at specific point in time. When there is no sample, we expect to measure a value equal to the amplitude of the applied sinusoidal magnetic field. The temperature dependence of the GaP detector is weak so we do not see any change in the response when cooling (see Fig. S6c). As described in Supplementary Information S5, the magnetic field is extracted as the difference between the polarization rotation measured when the polarity of the sinusoidal wave is positive minus the one measured when the polarity is negative. Fig. S6d schematically shows the evolution of the magnetic field upon cooling our sample across $T_c$. Initially, when the sample is above $T_c$, the magnetic field measured is almost equal to the applied magnetic field due to the higher resistivity of the thin film in the normal state. Upon cooling across $T_c$, as described above, the superconductor begins to partially shield the time varying applied field. In this temperature regime close to $T_c$, the shielding is not complete due to residual dissipation and to the low critical current density of the superconductor. Upon further cooling to lower temperatures the shielding improves, as the superfluid density increases, and the superconductor behaves like a perfect metal in which the eddy currents induced by the applied field become persistent due to the absence of dissipation. In this regime of perfect screening, the field on top of the sample is constant and our differential measurement yields a value of the magnetic field close to zero above the center of the superconductor as shown on the left side of Fig. 2b in the Main Text.

## S7. Field Cooled vs. Zero-Field Cooled Magnetization

When cooling a superconductor across $T_c$ in a magnetic field the Meissner effect takes place and the sample expels the magnetic field from its volume. In the presence of non-superconducting defects the field is trapped into these regions of the superconductor leading to a reduced magnetization[8,9] (see Fig. S7.1). The $YBa_2Cu_3O_7$ thin film used has a relatively high concentration of defects (see Supplementary Information S1). This leads to a great difference between the screening susceptibility (zero field cooled, see Supplementary Information S8) and Meissner effect (field cooled) responses of our thin film[8].

We used magnetostatic calculations to assess the effect of flux trapping defects on the Meissner response in our experimental geometry. We modeled the superconducting thin film as a uniform medium with a magnetic permittivity close to zero. The presence of defect was simulated by introducing gaps (width of 150 nm) in the sample spaced 10 μm from each other. The changes in the magnetic field surrounding the sample were calculated using COMSOL Multiphysics to solve Maxwell's equations.

We calculated the measured magnetic signal in our geometry by integrating the z-component of the magnetic field in the volume of the GaP detector traversed by the probe. This has been done in two steps, first by integrating along the thickness (75 um) of the detector and then by convoluting the result with a gaussian envelope representing the probe spot size (50 μm FWHM). The results of this calculation are shown in Fig. S7.2b. The presence of discontinuities, mimicking the effects of defects, strongly reduces the field cooled magnetic response in our configuration and is not spatially resolved, due to the spatial average across the thickness of the GaP detector and to the smoothing effect from the finite size of the probe pulse.

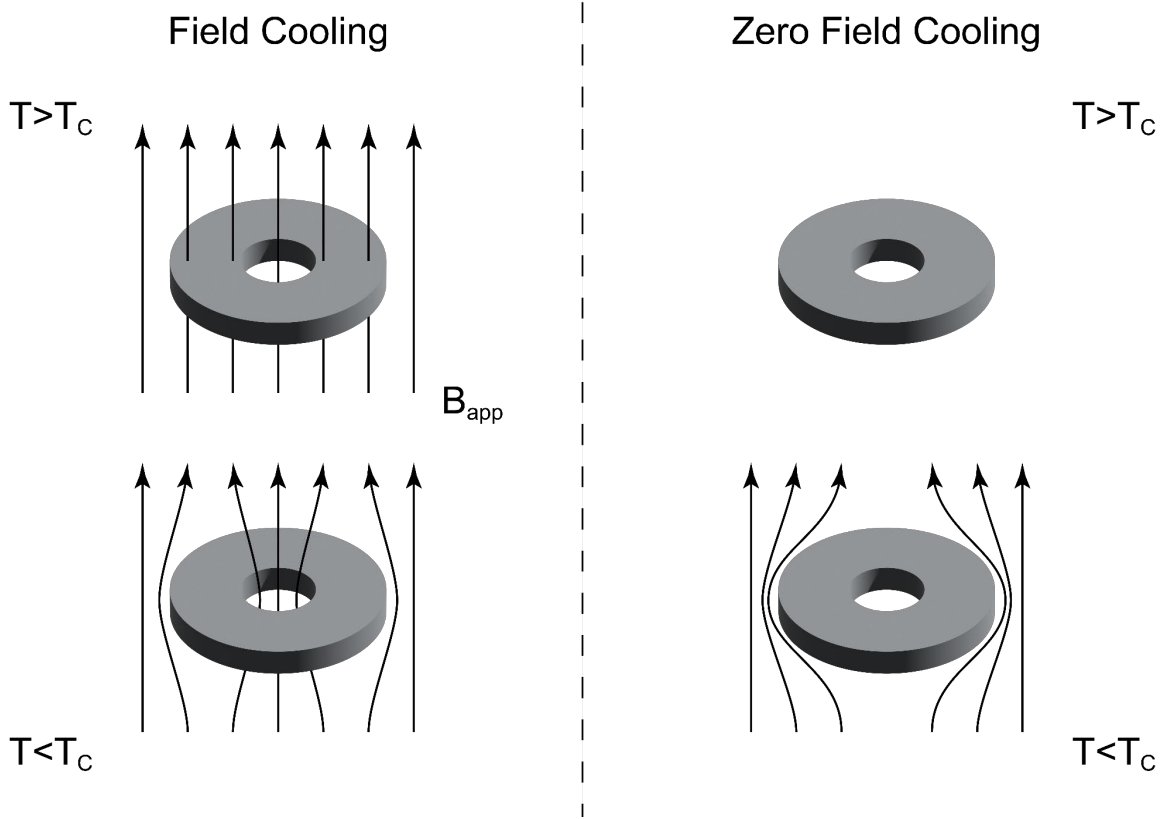

**Figure S7.1.** | Sketch representing how an applied magnetic field is modified by a superconducting sample with a hole, representing a spatial defect. When the sample is cooled in a magnetic field (left panel) the field is expelled from the sample and trapped into the hole, reducing the net magnetization of the sample compared to a defect free sample. Conversely, when the sample is first cooled below $T_c$ and then a magnetic field is applied (right panel), the field is excluded from the whole sample volume, no matter the presence of defects.

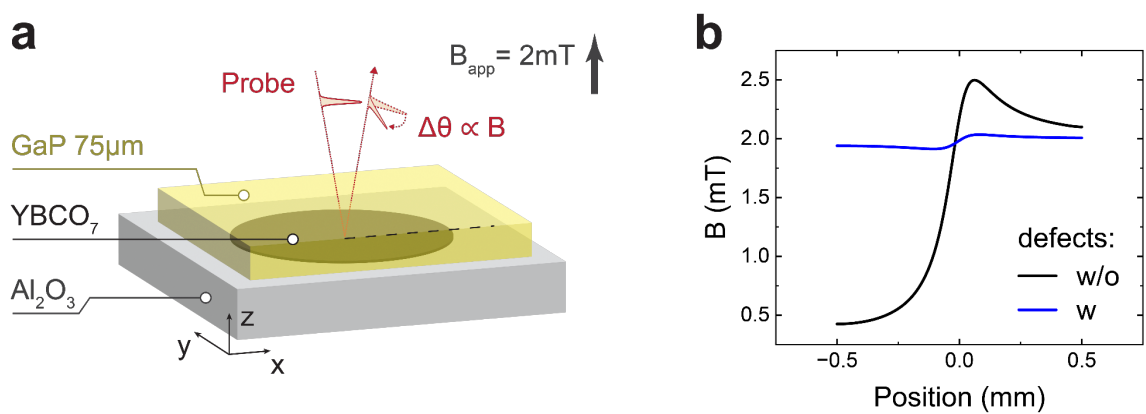

**Figure S7.2.** | **(a)** Geometry of the experiment. The black dashed line represents the direction of the spatial scan shown in b. **(b)** Simulated magnetic field spatial dependence across the edge of a perfectly diamagnetic thin disc with radius of 500 μm in a 2 mT externally applied field, with ("w") and without ("w/o") spatial defects measured in GaP.

This justifies our choice of resorting on zero-field cooled protocol in our measurements (see Supplementary Information S6), to maximize the diamagnetic response of the superconductor and to decrease the impact coming from the presence of defects.

In zero-field cooled conditions, the magnetic field exclusion originates from screening currents counteracting the change of magnetic flux threading the sample when an external magnetic field is applied once the sample is cooled below $T_c$. We simulated this situation with finite element calculations using COMSOL Multiphysics in a 3D geometry with axial symmetry.

The changes in the magnetic field surrounding the sample were calculated taking into account the geometry of the experiment. The solution domain was defined as a cylindrical region with a radius and a height of 1 mm. The $YBa_2Cu_3O_7$ sample was modeled as a disc with a radius of 500 μm and a thickness of 150 nm with a high conductivity ($10^{11}$ S/m). The magnetic field shielding from the superconductor (discussed qualitatively in Supplementary Information S6) was modeled by ramping up a uniform applied magnetic field from 0 to 2 mT in 1 ns, along the direction perpendicular to the plane of the disc. The magnetically induced eddy currents in the simulation reproduced the effect of the superconducting screening currents measured in the experiment, having the same physical origin: Faraday law. On this short time scales, the decay of the eddy currents was negligible (due to the high conductivity of the sample) and the magnetic shielding was constant, as expected from superconducting persistent currents. The substrate of the $YBa_2Cu_3O_7$ was not included in the modelling due to its much lower conductivity. Furthermore, we did not include effects due to the finite critical current of the sample, since the magnetic field applied in the experiment (2mT) was always lower the first critical field of YBCO (~100 mT[10]). Even taking into account demagnetizing effects appropriate for our geometry, we expect to reach the critical current threshold only in a small region close to the edge of the disc[6,7]. Therefore, our simulation matches our experimental conditions well and are in good agreement with the experimental results (see Supplementary Information S8). The simulations reported in Fig. 1 of the Main Text have been achieved with this method.

## S8. Additional spatial dependencies at equilibrium

The datapoints in Fig. S8a-b show respectively the z component of the magnetic field measured in the GaP and in the Bi:LIGG detectors as a function of the distance from the center of the YBa$_2$Cu$_3$O$_7$ disc, which is kept below $T_c$. The data clearly show a spatially dependent shielding of the applied magnetic field. We simulate this shielding with the zero-field cooled method described in Supplementary Information S7, first calculating the equilibrium magnetic field distribution in space generated by the YBa$_2$Cu$_3$O$_7$ sample (Fig. S9.3a) and then convoluting the averaged magnetic field across the thickness of each detector with a gaussian profile (50 µm FWHM) representing the probe beam. The distance of the detector from the superconductor used in the simulations is 45 µm for the GaP and 5 µm for the Bi:LIGG. Both these values are compatible with the geometry of the experiment. The dotted curves in Fig. S8 show the results of this simulation, which are in good agreement with the experimental data.

By comparing measured data and simulations we conclude that the spatial resolution of the GaP detector is limited by its thickness (75 µm) while for the Bi:LIGG detector the probe size (50 µm FWHM) is the limiting factor. This is clear if we assume a point like probe beam and compare again simulations (not shown in this case) and data: for the GaP the agreement is basically unchanged while for the Bi:LIGG the spatial features simulated are sharper than the ones measured. The Bi:LIGG detector measures a much larger enhancement of the magnetic field at the edge of the superconductor compared to the GaP because it is thinner and closer to the superconductor than the GaP.

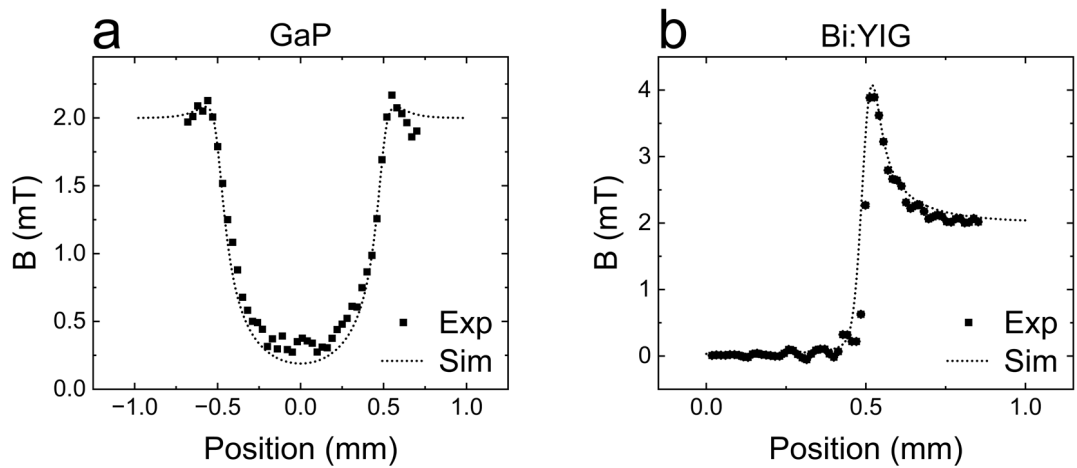

**Figure S8.** | Measured (datapoints) and simulated (dotted lines) z component of the magnetic field measured while scanning the position of the probe beam across the YBa$_2$Cu$_3$O$_7$ disc with **(a)** the GaP detector and **(b)** the Bi:LIGG sample.

Moreover, due to the spatial sensitivity of Bi:LIGG, we can put an upper limit to the region in the sample where the critical current is reached. In fact, the magnetic field can penetrate the volume of the superconductor in the area where a critical current is flowing[6,12]. From Fig. S8b below we can see that already less than 150 µm away from the edge of the sample the externally applied magnetic field is completely shielded. 100 μm is an upper limit, given the fact that the we do not correct for the finite size of the probe spot, which blurs the spatial resolution. This justifies the approximated model, which does not take into account the effect of critical currents, employed here and in Supplementary Information S7 and S9 to simulate the superconducting shielding.

## S9. Modeling the Ultrafast Magnetic Step

To gain deeper insight into the generation of ultrafast magnetic field steps, we started by assuming that the essential physics of this phenomenon can be capture with a simple circuit model characterized by an inductor (with inductance L) and a resistor (with resistance R). Indeed, the data shown in Fig. 2 of the Main Text can be fitted well by a single exponential decay function of the form (see Fig. S9.1):

$$y(t) = \begin{cases} y_0, & t < t_0 \\ y_0 + A \cdot \left(1 - e^{-\frac{t-t_0}{\tau}}\right), & t \geq t_0 \end{cases}$$

representing the typical response to a voltage step of an L/R circuit.

In this formula: t is the independent variable representing the time-delay, $y_0$ is the baseline at negative time delays accounting for the magnetic flux trapped at $t<t_0$, A is the amplitude of the magnetic step, $t_0$ is the arrival time of the quench beam and τ is the characteristic rise-time of the magnetic step. The value of ΔB shown in Fig. S9.1 is achieved by subtracting to the magnetic field measured the fitted value of $y_0$. The value of τ extracted from the fit is roughly 1.3 ps. In this picture the quench pulse creates a sudden increase of the resistivity of the disc, with the current decreasing on time scales dictated by the inductance and resistance of the photo-excited state. We proceeded by estimating the L/R time constant in our geometry, to more quantitatively validate this simple circuit model.

We used again COMSOL to extract the resistance and the inductance of a thin disk with a conductivity of $10^6$ S/m, a value similar to the normal state in-plane conductivity of YBCO at 100K[11]. We take into account that the current mainly flows at the edges of the disc[12] by modeling our sample as a ring with an internal radius equal to 70% of the external radius and biasing it with a uniform current distribution to extract L and R. Fig. S9.2 shows the linear dependence of the time constant of the system (equal to L/R) as a function of the external radius (r) of the ring. When r is set to 500 μm, the time constant extracted is 9.2 ps. This value is roughly 7 times bigger than what measured in the experiment (~1.3 ps).

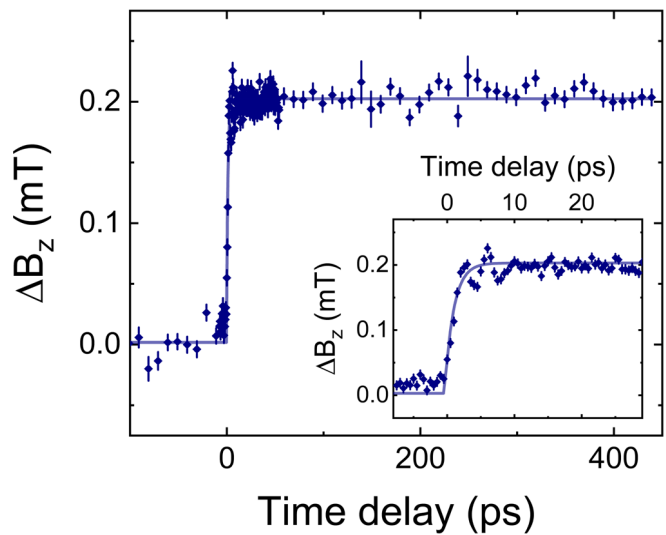

**Figure S9.1.** | Fit of the quench-induced changes in the *z*-component of the local magnetic field, $\Delta B_z$, measured above the centre of the disc as a function of pump-probe delay. The diamonds represent experimental datapoints and the solid line is the fit with a rise and single time constant exponential decay (see above). The inset shows a zoom-in at short time delays around time zero. The value of the measured magnetic field at negative time delays is equal to 0.45 mT due to the trapped flux in the superconductor between consecutive pump pulses (see Supplementary Information S13). The error bars denote the standard error of the mean.

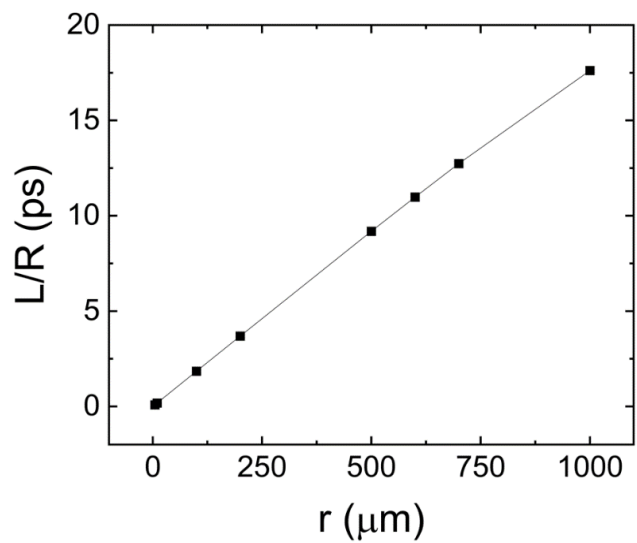

**Figure S9.2.** | Simulated time constant (L/R) of a resistive ring with conductivity 1 MS/m, thickness 150 nm and ratio internal to external radius fixed to 0.7, versus size of the external radius (r).

The biggest uncertainty in our model comes from the value used for the conductivity of the photo-excited state. Another factor that makes this estimate not completely accurate is the fact that we use a uniform current distribution to bias the sample and not the real spatial distribution of the current in the sample before the quench beam hits it. These two aspects can explain the discrepancy between experiment and simulations. However not perfectly correct, this model has still the advantage of being simple and of highlighting that reducing the diameter of the superconducting disc is the path forward to achieve faster switching times.

To further check that the magnetic dynamics of the magnetic step, we performed a time dependent finite element calculation using COMSOL Multiphysics, similar to the one discussed in Supplementary Information S7 to model zero-field cooled shielding currents and adding a sudden quench in conductivity to reproduce the effect of the quench pulse. The solution domain was defined as a cylindrical region with a radius and a height of 10 µm. The $YBa_2Cu_3O_7$ sample was modeled as a disc with a radius of 5 µm and a thickness of 150 nm with a high conductivity ($10^{11}$ S/m). We had to simulate a sample with a smaller lateral dimension than the one in our experiment due to the high computational cost of the simulation. Also here, the superconductive magnetic field shielding was modeled by ramping up a uniform applied magnetic field from 0 to 2 mT in 1 ns along the direction perpendicular to the plane of the disc, as discussed in Supplementary Information S7. We simulated the ultrafast magnetic step by quenching the conductivity of the thin disc to $10^6$ S/m, which is the conductivity assumed above for the photo-excited state. This is equivalent to assuming that in the experiment all the superconducting carriers are turned into normal carriers by the pump pulse. The results we achieved are shown in Fig. S9.3. Before the conductivity quench the equilibrium distribution of the magnetic field (Fig. S9.3a) is similar to what reported in Fig. 2 of the Main Text, with the difference that the spatial features here are sharper because we were not integrating across the thickness of the GaP detector (see Supplementary Information S8). Fig. S9.3b shows the dynamics of the magnetic field in the center of the disc after the conductivity quench. The shielding disappears on ultrafast time scale: after roughly 0.3 ps the magnetic field reaches the value of the external magnetic field. Also in this simulation, a single exponential fit (see equation above) captures accurately the time dependence shown in Fig. S9.3b. This fit yields a time constant of ~ 0.08 ps which is in very good agreement with the L/R time constant of 0.088 ps calculated above when setting r to 5 µm (see Fig. S9.2). This result further validates the simple circuit model we used above and consequently also its predictive value.

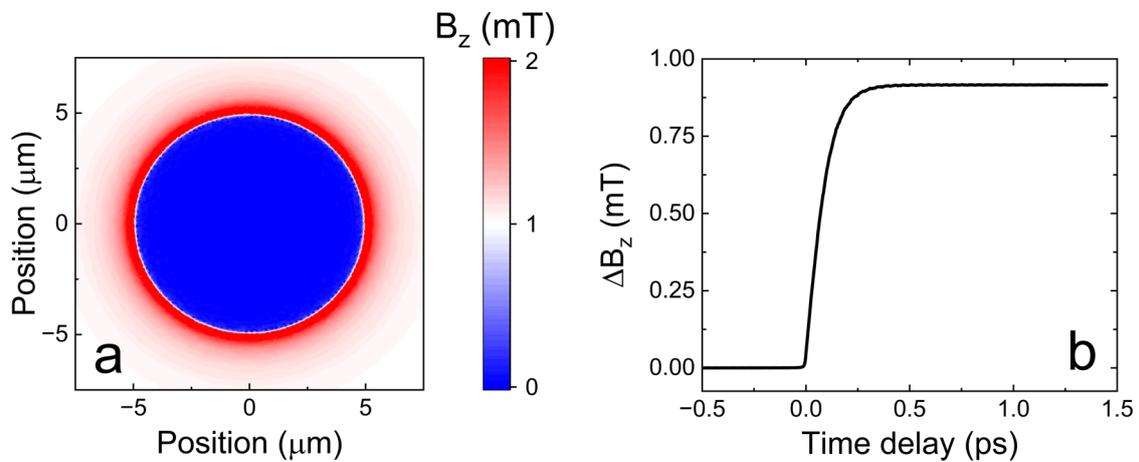

**Figure S9.3.** | **(a)** Simulated z component of the magnetic field ($B_z$) above the centre of the superconductor at negative time delays. **(b)** Change of $B_z$ in time after the conductivity quench.

## S10. Imaging System

Fig. S10 shows a schematic representation of the imaging setup used to excite the superconducting $YBa_2Cu_3O_7$ disc with a flat top beam matching exactly its shape.

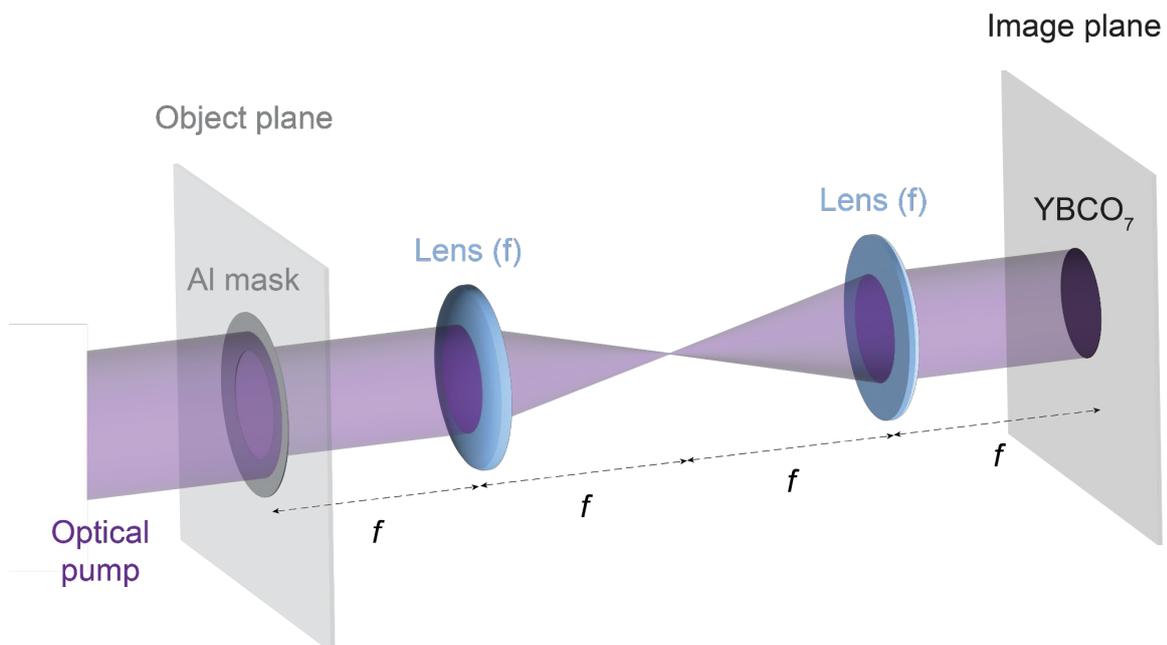

**Figure S10.** | Schematic drawing of the 4f imaging system used to selectively pump only the $YBa_2Cu_3O_7$ disc pattern.

A metallic mask is used to shape the pump beam. This mask is produced depositing a 300 nm thick aluminum thin film on a glass substrate, transparent to the 400 nm optical pump pulse. We then wet etched a disc shape on the Al film using a laser lithography process based on a AZ1512 photoresist mask. After exposure and development, the sample was wet etched using a commercial etchant solution for Al (TechniEtch Al-80 from MicroChemicals GmbH). After etching, the residual photoresist was removed using acetone and isopropanol. The optical pump, propagating trough the disc patterned into the Al film, was then imaged onto the sample using a 4f optical system. As the pump matches exactly the shape of the opaque $YBa_2Cu_3O_7$ sample, the GaP detector (Fig. 2) and the Bi:LIGG sample (Fig. 3) are fully shielded from the pump pulses. This aspect is important to avoid a possible undesired crosstalk between pump and probe (see also Supplementary Information S11).

## S11. Absence of Pump-induced Magnetic Field Changes Above $T_c$

We repeated the experiments in Fig. 2 and 3 of the Main Text at 100 K, above the superconducting transition temperature $T_c$ of the $YBa_2Cu_3O_7$ sample. The results of this measurement for both the GaP detector and the Bi:LIGG sample are indicated by the black dots in Fig. S11.1 and S11.2 respectively. The blue diamonds show the results of the corresponding measurements done in the same conditions below $T_c$. It is evident that the magnetic response disappears above $T_c$, confirming the magnetic nature of the signal measured and ruling out possible spurious pump-probe effects induced by the optical pump.

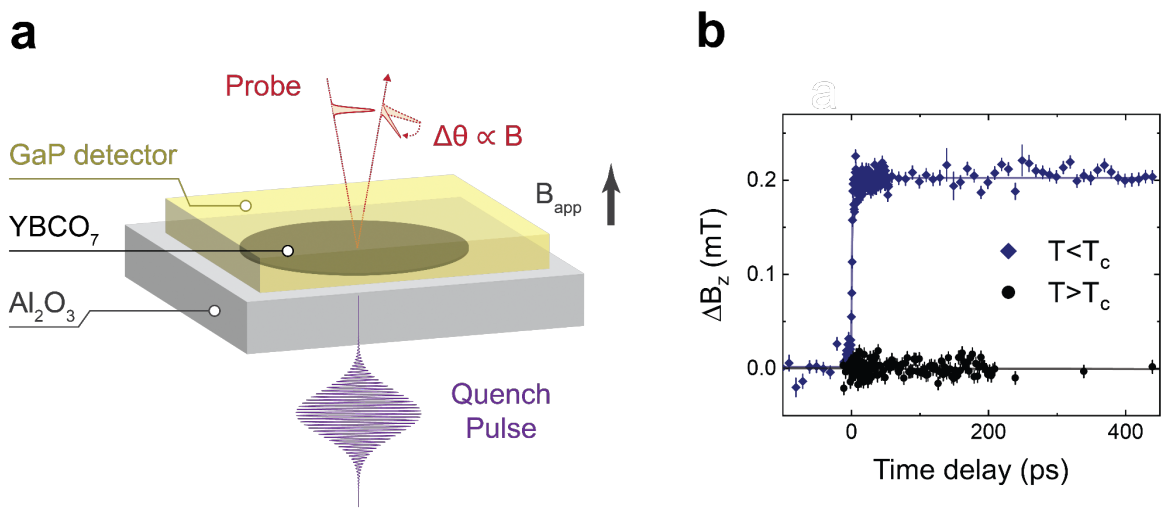

**Figure S11.1 | (a)** Geometry of the experiment. **(b)** Measured magnetic field above the centre of the superconducting disc below $T_c$ at 55 K (blue diamonds) and above $T_c$ at 100 K (black circles) in the GaP detector.

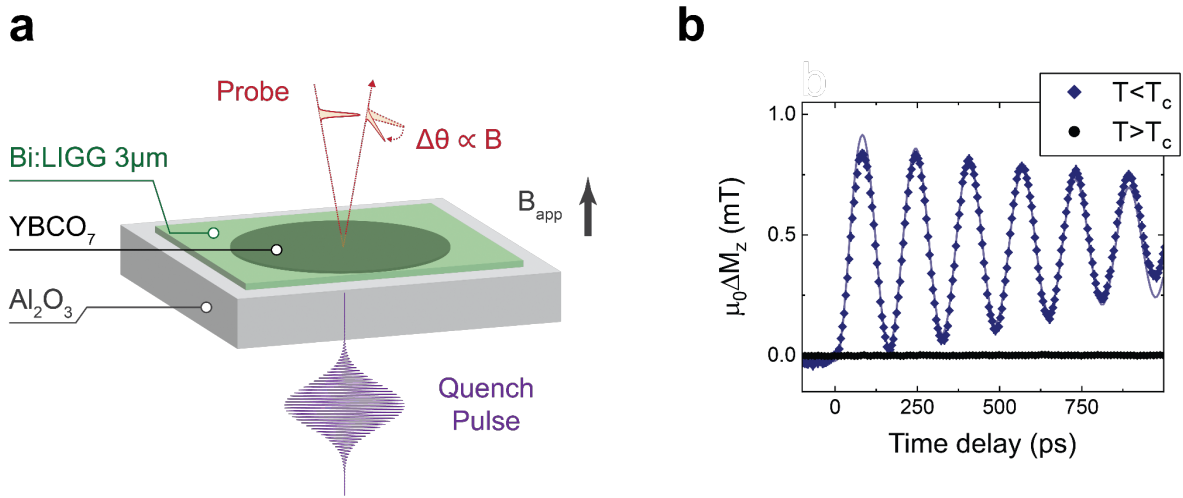

**Figure S11.2 | (a)** Geometry of the experiment. **(b)** Measured magnetic field above the centre of the superconducting disc below $T_c$ at 55 K (blue diamonds) and above $T_c$ at 100 K (black circles) in Bi:LIGG.

## S12. Fluence dependence of the Ultrafast Magnetic Step

Fig. S12 shows the measured magnetic steps for different values of incident pump fluence.

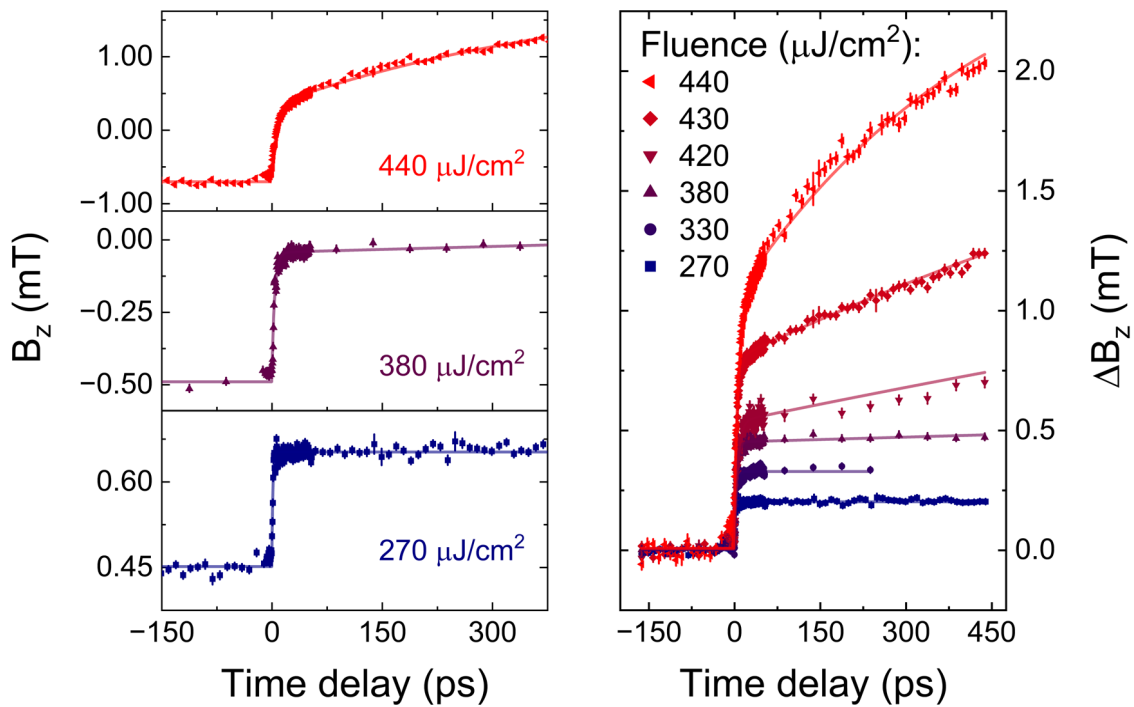

**Figure S12.** | Experimental data showing the measured magnetic field for three selected fluences of the quench beam (left panel) and the changes of the magnetic field for six different quench fluences (right panel) versus pump probe delay. These data have been measured using a GaP magneto-optic detector in the configuration shown in Fig. 2 of the Main Text.

The curve measured at the lowest fluence (270 µJ/cm²) is the same shown in Fig. 2 of the Main Text. As we progressively increase the fluence of the pump, the amplitude of the ultrafast magnetic step also increases. At a fluence of ~400 µJ/cm² the step suddenly changes its behavior and an additional, slower time scale appears. As discussed in Supplementary Information S13, the curves taken at higher fluence show the presence of trapped flux with opposite polarity at negative time delays. This is visible in the negative pump-probe time delay region of the first two graphs (440 and 380 µJ/cm²) in the left-hand-side panel of Fig. S12.

## S13. Pump Induced Magnetic Field Dynamics at Long Time Scales

Fig. 2 of the Main Text shows the ultrafast magnetic field changes induced by the quench pulse. Here instead, we want to qualitatively discuss the evolution of the absolute value of the magnetic field, determined by long-lasting thermal effects induced by the optical quench pulse and by slow magnetic flux dynamics. We start discussing in an ideal case the evolution of the magnetic field in the superconductor, assuming that the quench beam has enough energy to disrupt entirely the superconductive shielding (Fig. S13.1). We then model the magnetic field dynamics when the quench beam disrupts superconductivity only partially and after factoring in the contributions due to our measurement geometry (Fig. S13.2 and S13.3). This second case is relevant to the measurement shown in the Main Text.

Before the first quench pulse arrives, the sample is kept below $T_c$ and the magnetic field above the center of the superconductor is equal to zero as discussed in Supplementary Information S6 (see Fig. S13.1c showing that the magnetic field in the superconductor is zero before the first quench pulse arrives). When the first quench pulse hits the sample the magnetic field above the center of the superconductor suddenly reaches the value of the external magnetic field creating an ultrafast magnetic field step. The photo-excited sample will then thermalize and stay at a temperature higher than $T_c$ until the heat is dissipated through the Al$_2$O$_3$ substrate. During this transient the magnetic field is equal to the value of the externally applied magnetic field since there are no superconducting currents available for the shielding above $T_c$. The superconductor then recovers in an applied magnetic field, which is trapped in the defects of the sample (see Supplementary Information S7). At this point superconducting screening currents build up to shield any further change of the applied field and keep the value of the magnetic flux threading the sample constant (as prescribed by Faraday law), until the next pump pulse disrupts superconductivity again. At this point the process described above repeats and the magnetic field in the sample reaches again the value of the externally applied field on ultrafast timescales. Fig. S13.1 schematically depicts these dynamics and highlights the sign of the measured magnetic signal at negative (left panel) and positive (right panel) time delays. At negative delays, i.e., right before a quench pulse hits the sample, the trapped magnetic flux has a polarity opposite to the one of the applied field. This is the reason why the measured magnetic field at negative pump-probe delays *can* be negative in our measurements, as shown in Fig. S12. At positive time delays the value of the magnetic field measured in the vicinity of the sample is positive and close to the value of the external applied field as observed at long pump probe delays in the data taken at the highest fluence in Fig. S12.

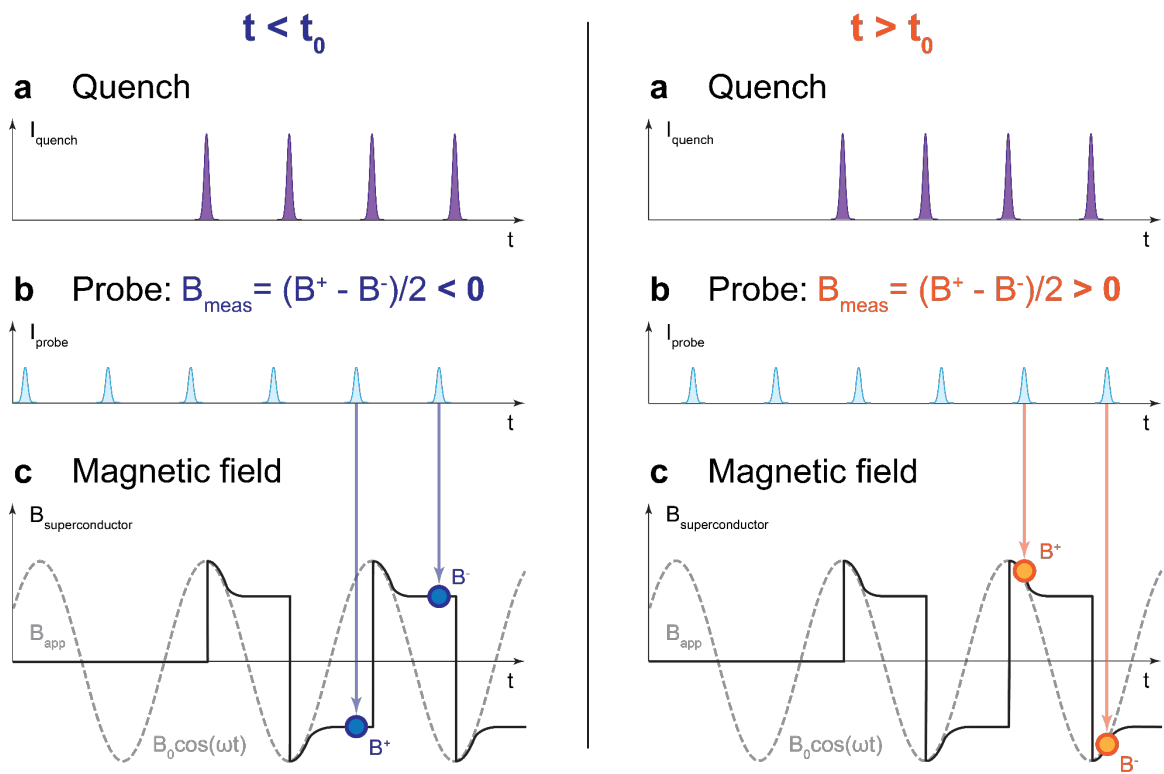

**Figure S13.1. | (a)** Arrival time of the quench pulses disrupting superconductivity in the sample (same in both the left and the right panels). **(b)** Probe pulses sampling the magnetic field. In the left panel the probe pulses arrive *before* the arrival time of the respective quench pulses ($t_0$) and the magnetic field measured is negative due to the presence of trapped flux with a polarity opposite to the externally applied field. In the right panel the probe pulses arrive *after* the respective quench pulses ($t>t_0$) and the magnetic field measured is positive and close to the value of the external

applied field. **(c)** External applied magnetic field (dashed grey line) and magnetic field in the center of the superconductor (solid black curve).

We note that the presence of trapped magnetic flux with polarity opposite to the applied field, can also lead to a magnetic step having amplitude bigger (up to twice) than the amplitude of the external applied field (as hinted by the curves at highest fluence in Fig. S12). The data reported in the Main Text have been acquired in a condition in which the energy of the quench pulse is not sufficient to entirely disrupt the shielding currents in the superconductor. Moreover, to describe the signal, we also need to account for the finite distance from the superconducting sample at which the field is measured. In fact, the applied magnetic field at the detector is not completely shielded by the superconductor as schematically depicted in Fig. S13.2 (see also Supplementary Information S8). The combination of these two aspects gives rise to magnetic dynamics reported in Fig. S13.3. Here the magnetic field after the quench is not reaching the external field, giving rise to a magnetic step with lower amplitude, and the measured field at negative time delays has the same sign of the applied field because the trapped flux with negative polarity is offset by the applied field. This picture qualitatively describes the results shown in Fig. 2b of the Main Text.

Finally, we present in Fig. S13.4 a possible alternative measurement protocol to avoid the presence of trapped flux at negative time delays. Here we assume that the quench pulses deposit enough energy on the sample to completely disrupt superconductivity and we apply a piecewise sinusoidal external magnetic field as shown by the dashed line in Fig. S13.4c. In this way, we can use half of the quench pulses to "reset" the magnetic field of the superconductor to zero by quenching superconductivity in a zero applied magnetic field. The field is then measured with probe pulses arriving at half the repetition rate of the quench pulses (Fig. S13.4b). Thus, the field right above the superconductor at negative time delays is going to be zero.

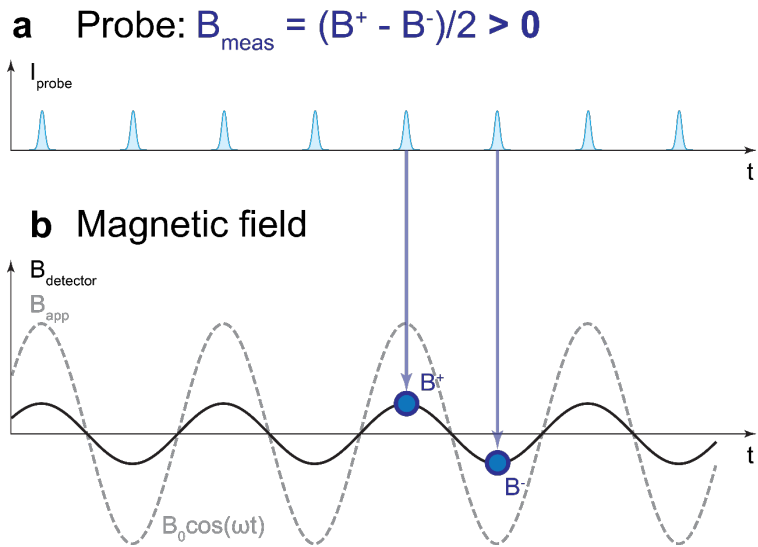

**Figure S13.2.** | **(a)** Probe pulses sampling the magnetic field. The magnetic field measured in the detector is positive because the shielding is not perfect at finite distances from the superconductor. **(b)** External applied magnetic field (dashed grey line) and magnetic field at the detector position above the center of the superconductor (solid black curve).

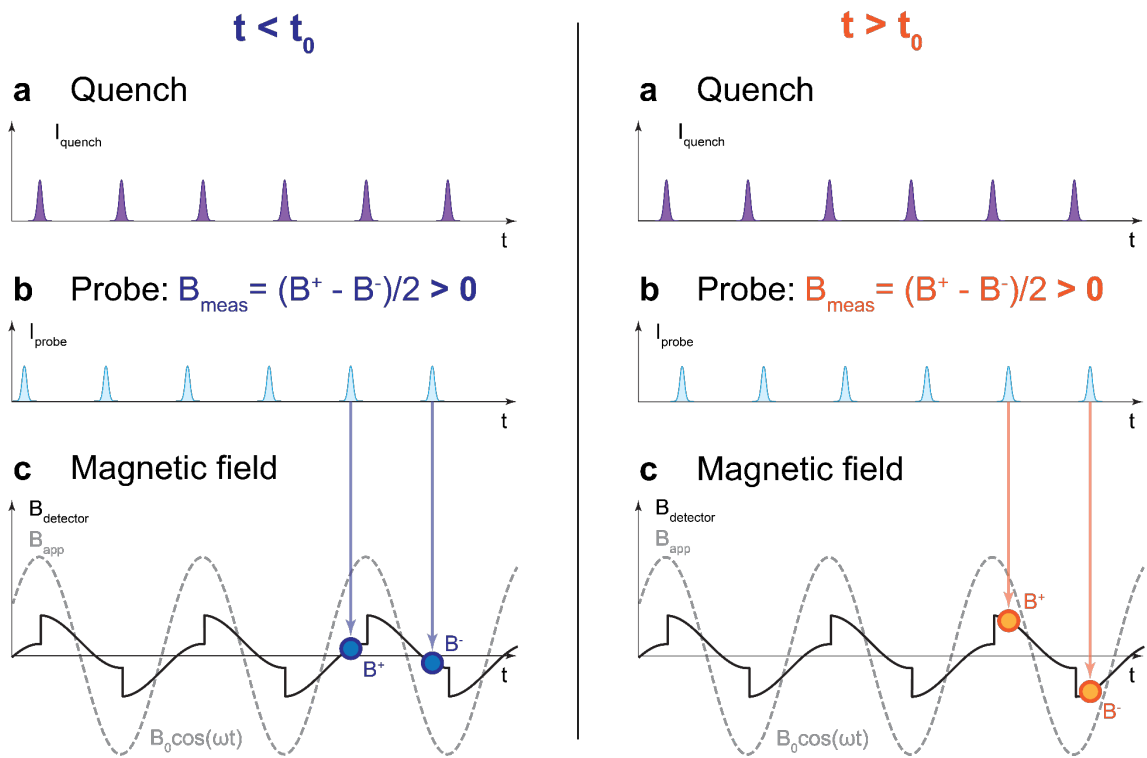

**Figure S13.3.** | **(a)** Arrival time of the low fluence quench pulses disrupting superconductivity in the sample (same in both the left and the right panels). **(b)** Probe pulses sampling the magnetic field. In the left panel the probe pulses arrive *before* the arrival time of the respective quench pulses ($t_0$). At the specific fluence shown here the magnetic field measured is still positive as the trapped flux with opposite polarity is offset by the applied field. In the right panel the probe pulses arrive *after* the respective quench pulses ($t>t_0$). The magnetic field measured is positive but not as big as the external field due to the only partial quench of superconductivity. **(c)** Externally applied magnetic field (dashed grey line) and magnetic field at the detector position above the center of the superconductor (solid black curve).

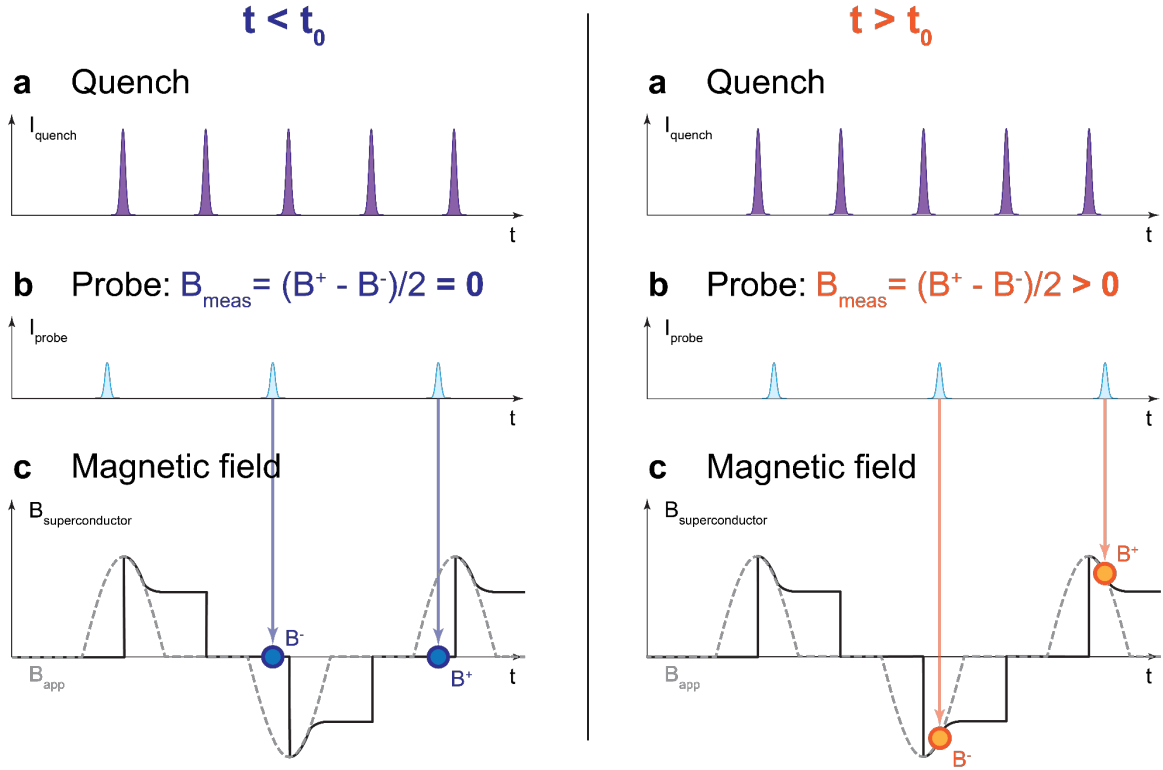

**Figure S13.4.** | **(a)** Arrival time of the quench pulses disrupting superconductivity in the sample (same in both the left and the right panels). **(b)** Probe pulses sampling the magnetic field at half the repetition rate of the quench pulse. In the left panel the probe pulses arrive *before* the arrival time of the respective quench pulses (t<$t_0$) and the magnetic field at negative time delays is equal to zero with this experimental scheme. In the right panel the probe pulses arrive *after* the respective quench pulses (t>$t_0$) and the magnetic field measured is positive and close to the value of the external applied field. **(c)** External applied magnetic field (dashed grey line) and magnetic field in the center of the superconductor (solid black curve).

## S14. Fitting the oscillations in Bi:LIGG

The magnetic oscillations shown in Fig. 3 of the Main Text were fitted with the following heuristic function capturing the damped magnetic oscillations observed in the data (see Fig. S14):

$$y(t) = y_0 + A \cdot (1 + \sin(2\pi f(t - t_0) + \varphi)) \cdot e^{-\frac{t-t_0}{\tau}} \cdot (1 +)/2$$

Where: t is the independent variable representing the time-delay, $y_0$ is the baseline at negative time delays accounting for the magnetic flux trapped at t<$t_0$, A is the amplitude of the magnetic step and of the magnetic oscillations, f is the frequency of the magnetic oscillations measured, $t_0$ is the arrival time of the quench beam, $\varphi$ is a phase term to account for the initial phase of the oscillations, τ is the characteristic damping time of the oscillations and σ is the width of the error function necessary to kill the exponential divergence of the fitting function at negative times while ensuring a smooth fit around time t=t₀ (with σ ∼ 0.1 ps). The value of ΔM shown in Fig. S14 is achieved by subtracting to the magnetization measured the fitted value of y₀. The values of f and of τ extracted from the fit are 6.17 GHz and 1.23 ns respectively.

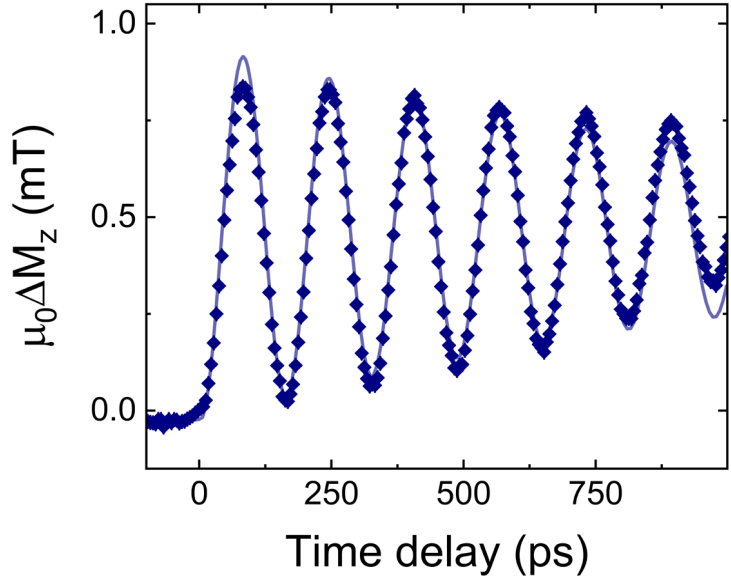

**Figure S14.** | Fit of the time dependent changes in the z-component of the magnetization ΔM$_z$ of Bi:LIGG, measured above the centre of the superconducting disk as a function of quench-probe delay. The value of the z-component of the magnetization at negative time delays is equal to -1.375 mT due to the presence of negative trapped flux before the pump pulse hits the sample (see Supplementary Information S13).

## S15. Micromagnetic Calculations

We used the COMSOL Micromagnetics Module developed by Dr. Weichao Yu (Institute for Nanoelectronic devices and Quantum computing, Fudan University) to perform the simulations shown in Fig. 4b of the Main Text. This package uses a finite element method to solve the Landau–Lifshitz–Gilbert (LLG) equation. We set the value of the saturation magnetization (M$_s$) of our sample to 175 kA/m and the value of the Gilbert damping to 0.017 in order to fit our data. We simplified the calculation by modeling our sample as a magnetized cubic monodomain of size 10 nm, analogous to a single macro-spin. To account for shape anisotropy, we artificially inserted an effective magnetic field orthogonal to the applied magnetic field, having amplitude M$_s$. This is because the Bi:LIGG thin film sample is characterized by a demagnetizing factor along the out of plane direction roughly equal to 1[13,14]. The out of plane direction is thus a hard axis for the magnetization that can be described by an energy term equal to $E_{shape} = \frac{1}{2}\mu_0 M^2 \cos^2(\theta)$, where $\theta$ is the angle between the film normal and the magnetization vector M (see for instance Blundell chapter 6 paragraph 7.2 [14]). When no magnetic field is applied the magnetization lies naturally in the plane of the film, while a perpendicular field cants the magnetization to the out of plane direction. For small out of plane rotations of the magnetization ($\theta \sim \frac{\pi}{2} + \varepsilon$, with $\varepsilon \ll 1$) the energy term $E_{shape}$ can be modeled by introducing an effective magnetic field lying in the plane of the film and pointing along the direction of the equilibrium magnetization with a modulus equal to the magnetization itself. This corresponds to an energy term: $E_{eff} = -M \cdot H_{eff} = -\mu_0 M^2 \cos(\frac{\pi}{2} - \theta)$, with $\theta$ defined as above. If we substitute $\theta \sim \frac{\pi}{2} + \varepsilon$ into $E_{eff} = -\mu_0 M^2 \cos(\varepsilon) \sim -\mu_0 M^2 + \frac{1}{2}\mu_0 M^2 \varepsilon^2$ and into $E_{shape} = \frac{1}{2}\mu_0 M^2 \cos^2(\frac{\pi}{2} + \varepsilon) = \frac{1}{2}\mu_0 M^2 \sin^2(\varepsilon) \sim \frac{1}{2}\mu_0 M^2 \varepsilon^2$ we see that, modulo a renormalization of the energy independent on the angle, the two energy terms have the same functional form and therefore yield the same force on the sample magnetization in the LLG equation. Finally, the small angle approximation invoked above is justified in our case since the amplitude of the perpendicular magnetic field applied (i.e., the ultrafast magnetic step) is at most 3 mT while $M_s$ = 175 kA/m corresponding to a field of roughly 220 mT. The angular deviation from the plane is therefore smaller than ~1°. In this way we demonstrated that it is justified to model shape anisotropy with a static effective magnetic field lying in the plane.

We then applied a perpendicular magnetic field with the same temporal profile of the magnetic step measured in GaP and we tracked the time evolution of the z component of the magnetization of the macro-spin representing the Bi:LIGG sample. Fig. 4b in the Main Text shows the results of this simulation when the magnetization dynamics is triggered by a magnetic step similar to the one reported in Fig. 2b. These simulation match qualitatively well the experimental results shown in Fig. 3b of the Main Text.

# References (Supplementary Information)